\documentclass[preprint, authoryear]{elsarticle}

\usepackage{amsmath,amssymb,amsfonts}
\usepackage[ruled]{algorithm}
\usepackage[noend]{algpseudocode}

\usepackage{graphicx}
\usepackage{textcomp}
\usepackage{xcolor}
\usepackage{subcaption}
\usepackage{tabularx}
\usepackage{booktabs}
\usepackage[hyphens]{url}
\usepackage{nicefrac}
\usepackage{natbib}
\usepackage{geometry}

\algnewcommand\Input{\item[\textbf{Input:}]}
\algnewcommand\Output{\item[\textbf{Output:}]}
\algnewcommand\Send[2]{\textbf{send }\texttt{#1}\ifthenelse{\equal{#2}{}}{()}{(#2)}}
\algnewcommand\Broadcast[2]{\textbf{broadcast }\texttt{#1}\ifthenelse{\equal{#2}{}}{()}{(#2)}}
\algnewcommand\Receive[2]{\textbf{receive }\texttt{#1}\ifthenelse{\equal{#2}{}}{()}{(#2)}}
\algnewcommand\Calls[2]{\texttt{#1}\ifthenelse{\equal{#2}{}}{()}{(#2)}}%

\algrenewtext{Procedure}[2]%
    {\algorithmicprocedure\ \texttt{#1}\ifthenelse{\equal{#2}{}}{}{(#2)}}%

\newtheorem{definition}{Definition}

\begin{document}
%\title{A problem-solving strategy for self-adaptation in multiagent systems}
\title{A cooperative strategy for diagnosing the root causes of quality requirement violations in multiagent systems}
% A Cooperative Strategy for Identifying and Solving Problem Causes
% A Cooperative Problem-Solving Strategy for Adaptive Components
% An Adaptive Strategy to Cooperatively Identify and Solve Problem Causes

\author[ufrgs]{Jo\~{a}o Faccin\corref{cor1}}
\ead{jgfaccin@inf.ufrgs.br}

\author[ufrgs]{Ingrid Nunes}
\ead{ingridnunes@inf.ufrgs.br}

\author[con]{Abdelwahab Hamou-Lhadj}
\ead{wahab.hamou-lhadj@concordia.ca}

\cortext[cor1]{Corresponding author}
\address[ufrgs]{Universidade Federal do Rio Grande do Sul, Porto Alegre, RS, Brazil}
\address[con]{Concordia University, Montreal, QC, Canada}

\begin{abstract}
Many modern software systems are built as a set of autonomous software components (also called agents) that collaborate with each other and are situated in an environment. To keep these multiagent systems operational under abnormal circumstances, it is crucial to make them resilient. Existing solutions are often centralised and rely on information manually provided by experts at design time, making such solutions rigid and limiting the autonomy and adaptability of the system. In this work, we propose a cooperative strategy focused on the identification of the root causes of quality requirement violations in multiagent systems. This strategy allows agents to cooperate with each other in order to identify whether these violations come from service providers, associated components, or the communication infrastructure. From this identification process, agents are able to adapt their behaviour in order to mitigate and solve existing abnormalities with the aim of normalising system operation. This strategy consists of an interaction protocol that, together with the proposed algorithms, allow agents playing the protocol roles to diagnose problems to be repaired. We evaluate our proposal with the implementation of a service-oriented system. The results demonstrate that our solution enables the correct identification of different sources of failures, favouring the selection of the most suitable actions to be taken to overcome abnormal situations.
\end{abstract}


% make the title area
\maketitle

% !TEX root = main.tex

\section{Introduction}\label{sec:introduction}

There is a wide range of software systems that are nowadays built as multiagent systems (MAS)~\citep{Jennings:01}. They are composed of distributed autonomous components that interact with each other and are situated in an environment. Examples of such systems include some of those used in unmanned air vehicles~\citep{Insaurralde:14}, autonomous vehicles~\citep{Huang:20}, and bike sharing systems~\citep{Dotterl:17}. Due to dynamic collaborations that emerge in these systems and its evolving environment, it is crucial to adopt techniques to make them resilient. That is, these systems must operate satisfying the required quality levels for as long as possible, resisting and recovering from abnormal situations~\citep{dobson:19, deSanctis:20}. This can be achieved by providing systems with self-adaptive~\citep{dobson:06} capabilities, which allow them to diagnose, remediate and, possibly, solve a given problem, which can be an unexpected behaviour or a degradation in the system performance. By uncovering the root causes of existing problems, it is possible to take corrective measures in order to remedy and solve them, thus preventing future failures and keeping the system operational~\citep{faccin:18a}. 

There are many approaches that are able to detect anomalous behaviour in software systems~\citep{chandola:09}. Although these can be adopted to identify that a service provided by an agent is abnormal, they are not sufficient to diagnose and solve the root causes of this abnormality in a MAS. To provide a service, an agent may rely on services provided by other agents, which in turn may use services from their neighbours. Consequently, problems that occur at runtime can be due to any service on the service chain or even in the communication between agents. Despite the approaches for anomaly detection are able to automatically raise system abnormalities, diagnosing their causes requires an understanding of the system behaviour and its environment. This task is typically carried out by experts through the manual inspection of execution-related information obtained from sources such as tracing and profiling tools~\citep{zhou:18}. This is usually impractical for large MAS with a high number of agents with many interconnections. In these scenarios, the large amount of data to be inspected can eventually mask the truly important information, making it difficult to identify the real issue. It is also a time-consuming task that can prevent a required immediate response in order to keep the system operating normally. This dependency on experts to diagnose problem causes goes against the increasing need for autonomous systems, which are expected to overcome and adapt to challenging situations without the need for human intervention~\citep{baresi:06}. In addition, existing techniques able to autonomously identify causal relationships from system execution has shown to be unable to efficiently handle distributed scenarios in a decentralised manner~\citep{maes:07, parida:18, wang:18}. 

In this paper, we provide a new approach in which agents autonomously diagnose, remedy, and repair the causes of quality requirement violations in MAS. Having the goal of adapting the system while keeping it operational, our decentralized solution consists of two main elements: (a) an interaction protocol that specifies roles and how they interact; and (b) a set of algorithms that define how agents behave in order to diagnose the problem to be repaired.
%In this paper, we address the problem of autonomously identifying problem causes at runtime with the aim of supporting the execution of a remedial problem-solving strategy. We consider a multiagent system in which an autonomous agent identifies an abnormal situation and seeks for the problem cause to be able to self-adapt the system and keep the system operational. Our proposed decentralised solution consists of an interaction protocol that specifies roles and how they interact. It also provides a set of algorithms to allow agents playing these roles to diagnose the problem to be repaired.
Our evaluation considers a service-oriented system, in which we inject different types of failures. The results show that, with our approach, it is possible to diagnose problem causes and self-adapt the system so that it maintains its desired behaviour. In summary, the contributions of this work are the following:
\begin{itemize}
 \item the specification of a strategy comprising an interaction protocol and a set of roles, which describes how agents can cooperate in order to diagnose the causes of quality requirement violations at runtime through the exchange of relevant information;
 \item the specification of how agents behave when playing the different roles specified in our protocol; and
 \item a practical evaluation of our proposal with the development of a service-based infrastructure.
\end{itemize}
%Next, we introduce definitions and detail our problem. Section~\ref{sec:solution} presents our solution, which is evaluated in Section~\ref{sec:evaluation}. Finally, related work is presented in Section~\ref{sec:related-work}, while conclusions are presented in Section~\ref{sec:conclusion}.

% !TEX root = main.tex

\section{Problem and Definitions}\label{sec:problem}

A MAS is a collection of autonomous components (agents) that are situated in an environment and are able to interact and collaborate by consuming and providing sets of services~\citep{Jennings:01}. We assume that, to execute their tasks with the expected quality, the agents require the services they consume to satisfy a set of predefined quality requirements, which are expressed in terms of measurable features and their range of acceptable values. An example of a quality requirement expressed in natural language is: \emph{``no service may have response time greater than 10 ms.''} When a quality requirement is not satisfied, there is a quality requirement violation, hereafter referred to as an \textit{abnormality}.
Concepts associated with services and their quality are defined as follows.

\begin{definition}[Service]
   A service $s \in \mathcal{S}$ is an action that can be performed by an agent.
\end{definition}

\begin{definition}[Quality Feature]
  A quality feature $q \in \mathcal{Q}$ is a measurable property associated with a service.
\end{definition}

\begin{definition}[Quality Requirement]
 Quality requirement $\mathcal{R}: \mathcal{Q} \nrightarrow \mathcal{K}$ is a function that maps a quality feature $q \in \mathcal{Q}$ to a constraint $\mathcal{K}$, which denotes the requirements to which a $s$ must satisfy with respect to each $q$ for which $\mathcal{R}$ is defined. $\mathcal{K}$ is expressed with the following grammar.
 \begin{align*}
  \mathcal{K} &::= (\mathcal{K} \wedge \mathcal{K})~|~(\mathcal{K} \vee \mathcal{K})~|~(\neg \mathcal{K})~|~e\\
  e &::= (q~op~v)\\
  op &::=~>~|~\geq~|~<~|~\leq~|~=~|~\not=
 \end{align*} where $q$ is the mapped quality feature and $v$ is a value $v \in \mathbb{R}$. 
\end{definition}

Agents interact through the exchange of messages. Our protocol considers two types of messages. \emph{Request} messages are sent by agents in order to ask for services or particular information. \emph{Inform} messages, in turn, are sent in order to either reply to requests or provide information that was not previously required. Messages with no recipient specified are broadcast to every component within a system.

\begin{definition}[Message]
 A message is a tuple $\langle id_m, id_c, c_s, c_r, type, s, cont \rangle$, where $id_m$ is the identifier of the message; $id_c$ is the identifier of the conversation of which the message is part; $c_s$ is the message sender; $c_r$ is the message receiver; $type \in \{inform, request\}$ is the type of the message; $s$ is the service that may be associated with the message, if any; and $cont$ is the message content, which can be of any type.
\end{definition}

\begin{figure}
 \centering
 \begin{subfigure}[b]{0.4\linewidth}
  \centering
  \includegraphics[width=.85\textwidth]{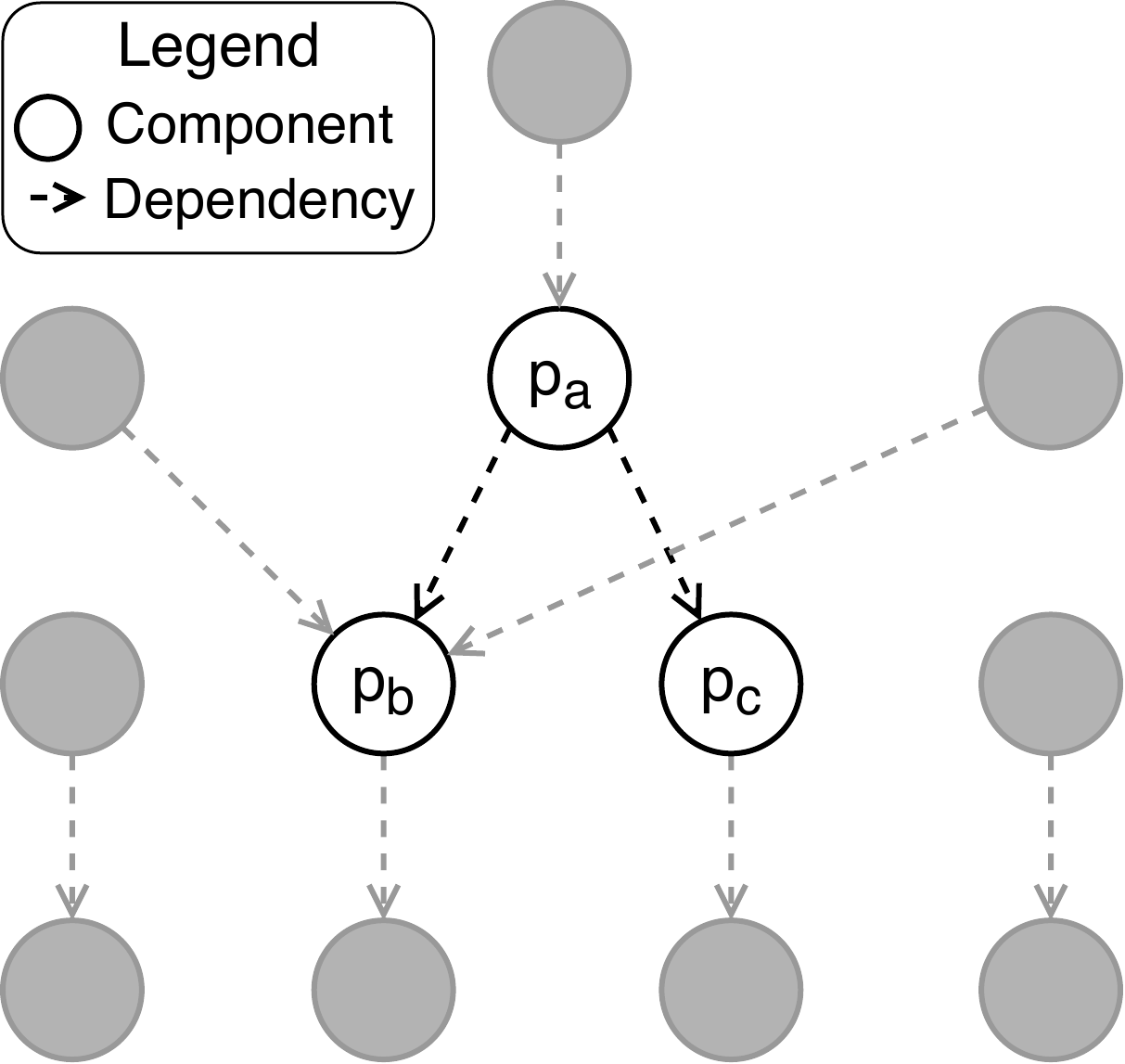}
  \caption{}
  \label{fig:example01}
 \end{subfigure}
 \begin{subfigure}[b]{0.4\linewidth}
  \centering
  \includegraphics[width=.85\textwidth]{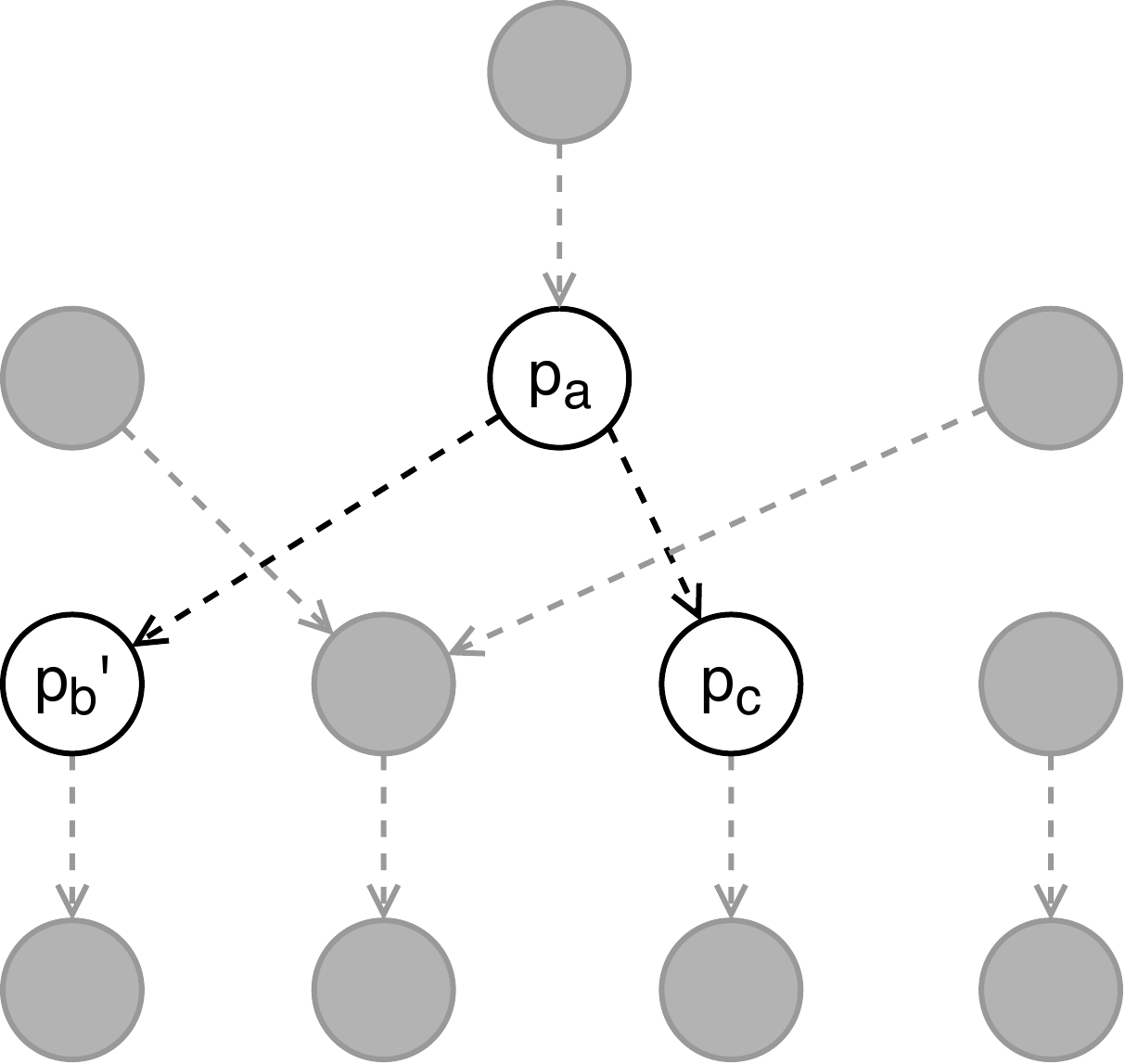}
  \caption{}
  \label{fig:example02}
 \end{subfigure}
 \caption{A MAS in which components interact with each other by consuming and providing services. (a) Agent $p_a$ relies on services $b$ and $c$ from agents $p_b$ and $p_c$, which, in turn, consume services from other agents. (b) Agent $p_a$ replaces service provider $p_b$ with ${p_b}'$.}
 \label{fig:example}
\end{figure}

Agents use services provided by other agents to make their own services available. As a result, when there is an abnormality, there are three possible sources of failure. An abnormality cause can be located at the agent itself (i.e.\ an \emph{internal cause}), at its service providers, or at their communication channel (both being \emph{external causes}). As an example, consider the system depicted in Figure~\ref{fig:example01} in which agents (represented as circles) interact by requesting and delivering services (dependencies between agents are represented as traced arrows from clients to their providers). To provide service $a$, component $p_a$ depends on services $b$ and $c$, consumed from $p_b$ and $p_c$, respectively. In order to deliver their own services, $p_b$ and $p_c$, in turn, rely on services from other agents. In this context, an abnormality presented by $p_a$ may not be necessarily caused by that component. Instead, its dependency on services from $p_b$ and $p_c$ indicates that these providers, or even those from which they consume services, could also be associated with the problem.

To diagnose the cause of an abnormality and be able to remedy and solve it, the system is required not only to recognise the existence of these component interactions, but also to evaluate which of them originated the abnormal behaviour. In a simple system, such as the one described above, it is possible to maintain a model that contains this causal information. For instance, if $p_a$ replaces service provider $p_b$ with ${p_b}'$ (Figure~\ref{fig:example02}), this new provider becomes a potential cause of further abnormalities presented by $p_a$. Consequently, this new dependency must be included in the existing causal model while the previous one must be removed from it as it is no longer a possible cause of abnormality. When we consider dynamic, large scale systems, however, maintaining such causal model in a centralised way is impracticable, either because of the amount of dependencies to be managed, or their dynamic evolution. Our goal is thus to autonomously diagnose the root cause of abnormalities and use this information to remedy and solve them in order to comply with predefined quality requirements.
%In this scenario, keeping inconsistent information may prevent the system from identifying the root cause of problems and, consequently, overcoming situations in which quality requirements are not met. Our problem can thus be stated as follows. \emph{How to enable system components to autonomously identify the root cause of abnormalities and use this information to remedy and solve them in order to comply with predefined quality requirements?}

% !TEX root = main.tex

\section{Cooperative Diagnosis and Solution of Problem Causes}\label{sec:solution}

To achieve our goal, we propose a strategy that allows agents to collaborate in order to diagnose and further remedy and solve the root cause of detected abnormalities. We achieve this by designing an interaction protocol that guides the interaction among agents and how they should behave when playing the different roles specified in our protocol. Next, we present an overview of our solution.

\subsection{Overview}\label{sec:solution:overview}

As mentioned in the previous section, the cause of an abnormality can be internal or external to an agent. In the latter case, it can be located at the level of service providers or the communication links. In this section, we present our solution that covers these scenarios to diagnose the origin of the detected abnormalities. It includes an interaction protocol and the specification of algorithms for agents to be able to play the protocol roles to diagnose abnormalities. 

\begin{figure*}[t]
 \centering
 \begin{subfigure}[b]{0.24\textwidth}
  \centering    
  \includegraphics[width=.9\textwidth]{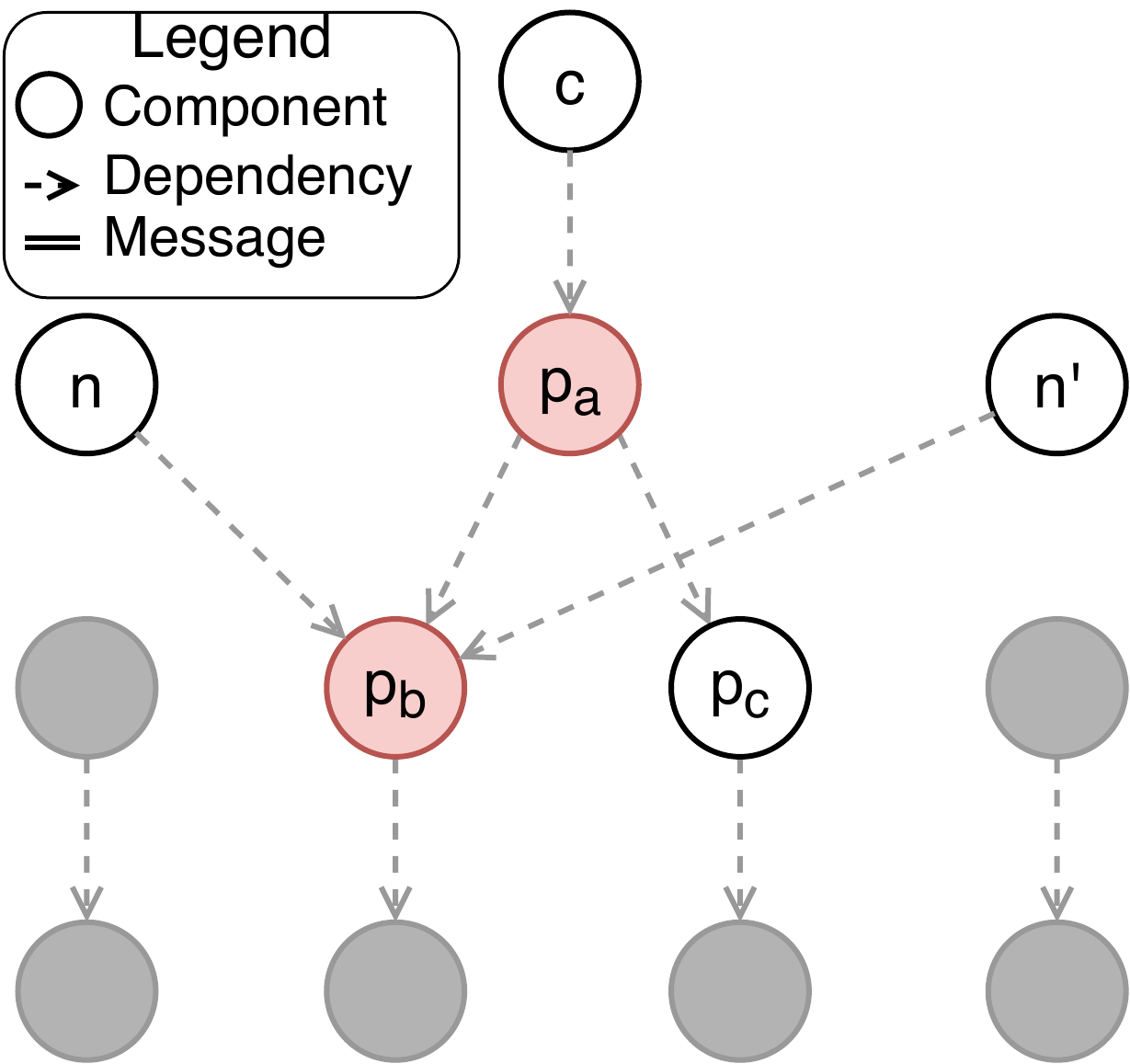}
  \caption{}
  \label{fig:overview01}
 \end{subfigure}
 \begin{subfigure}[b]{0.24\textwidth}
  \centering
  \includegraphics[width=.9\textwidth]{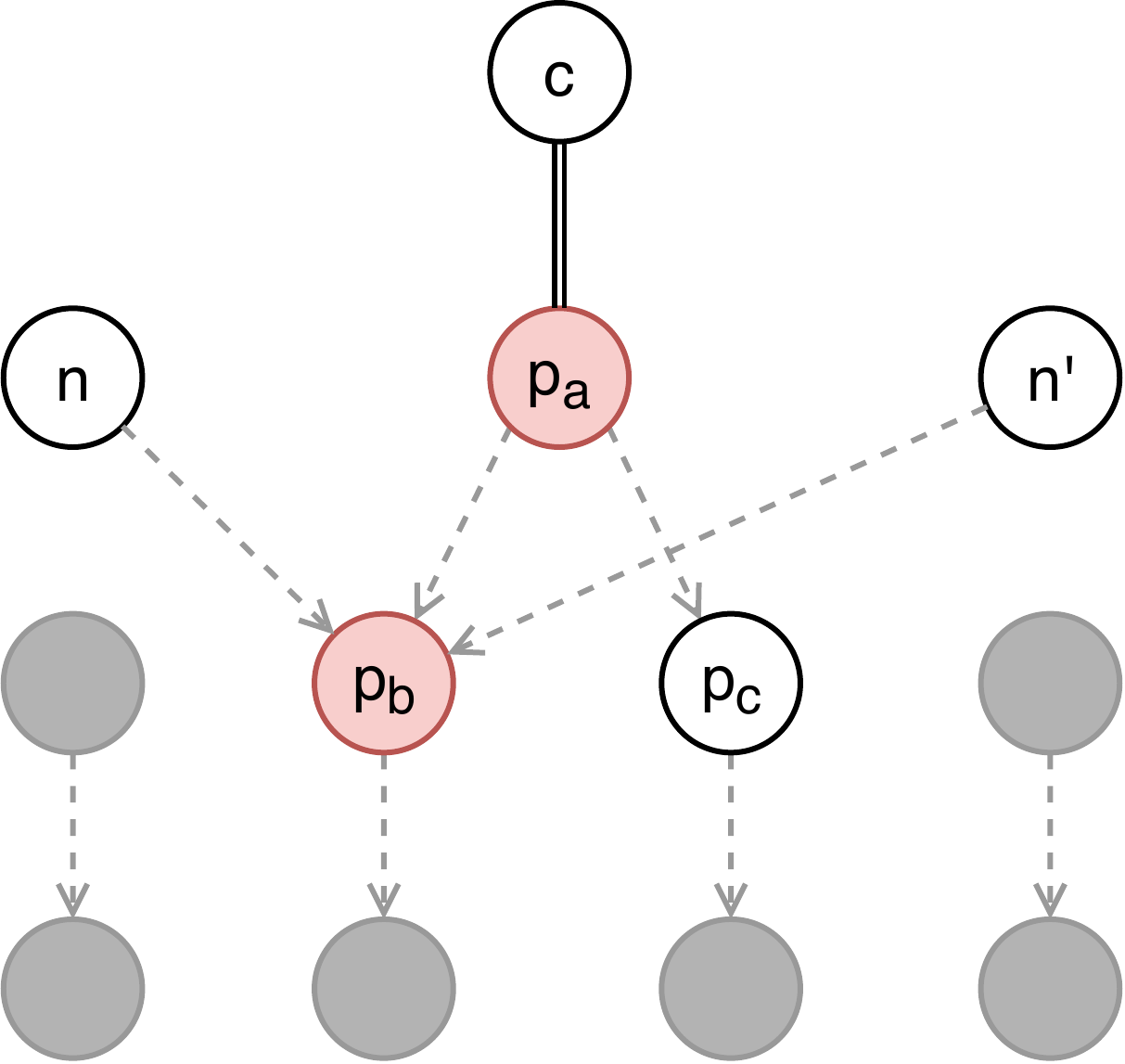}
  \caption{}
  \label{fig:overview02}
 \end{subfigure}
 \begin{subfigure}[b]{0.24\textwidth}
  \centering    
  \includegraphics[width=.9\textwidth]{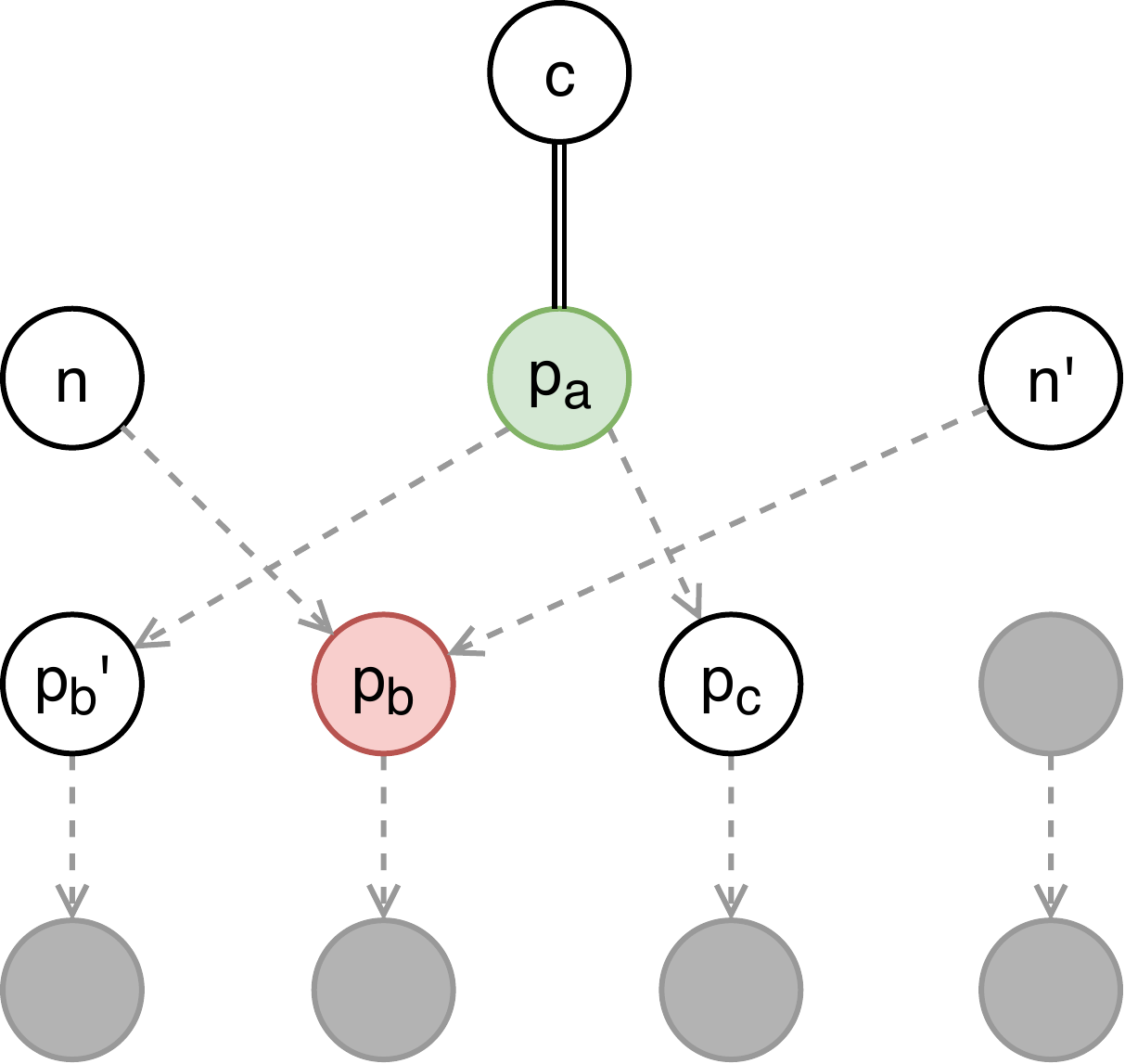}
  \caption{}
  \label{fig:overview03}
 \end{subfigure}
 \begin{subfigure}[b]{0.24\textwidth}
  \centering
  \includegraphics[width=.9\textwidth]{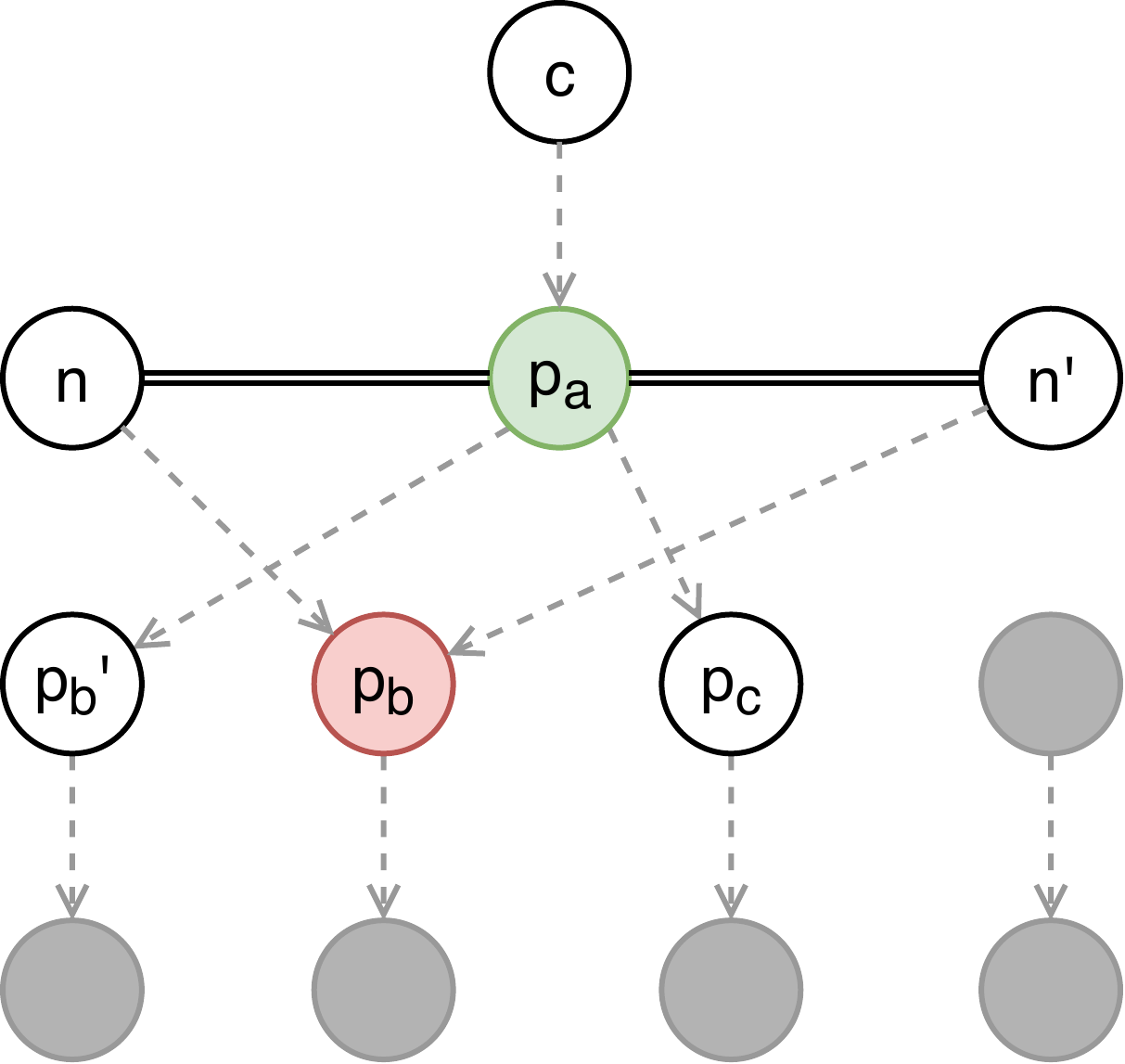}
  \caption{}
  \label{fig:overview04}
 \end{subfigure}
 \begin{subfigure}[b]{0.24\textwidth}
  \centering    
  \includegraphics[width=.9\textwidth]{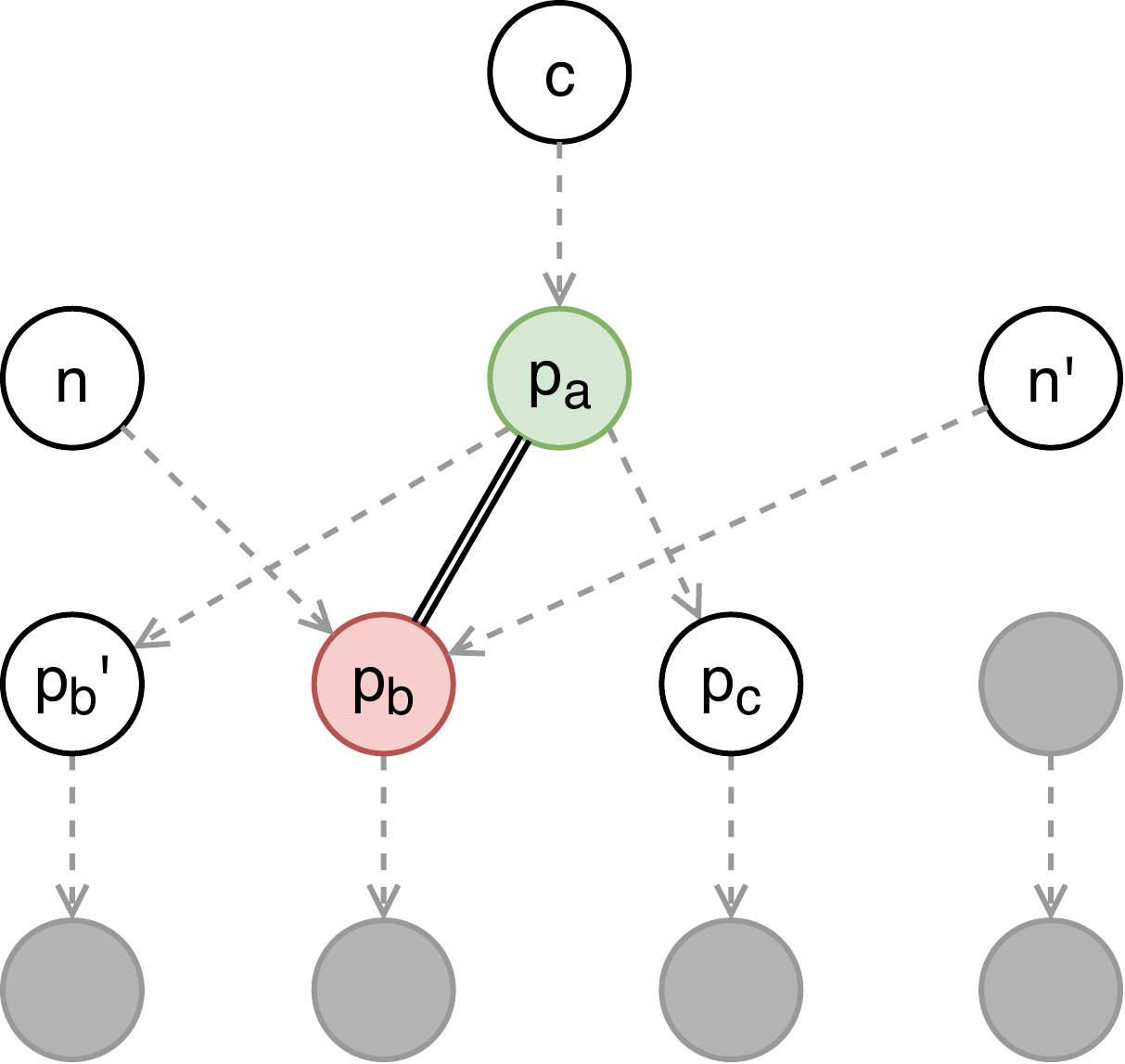}
  \caption{}
  \label{fig:overview05}
 \end{subfigure}
 \begin{subfigure}[b]{0.24\textwidth}
  \centering
  \includegraphics[width=.9\textwidth]{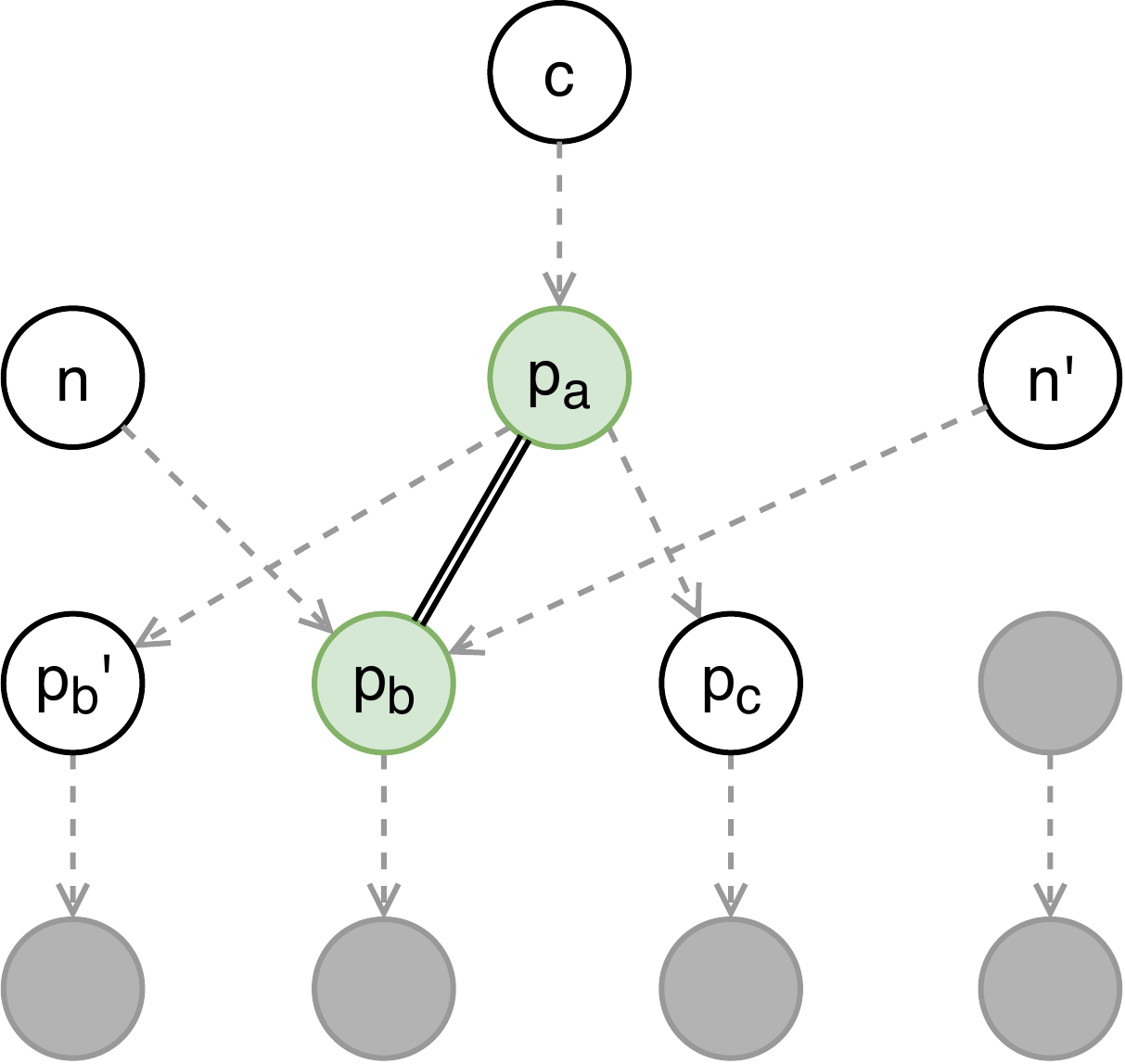}
  \caption{}
  \label{fig:overview06}
 \end{subfigure}
 \begin{subfigure}[b]{0.24\textwidth}
  \centering
  \includegraphics[width=.9\textwidth]{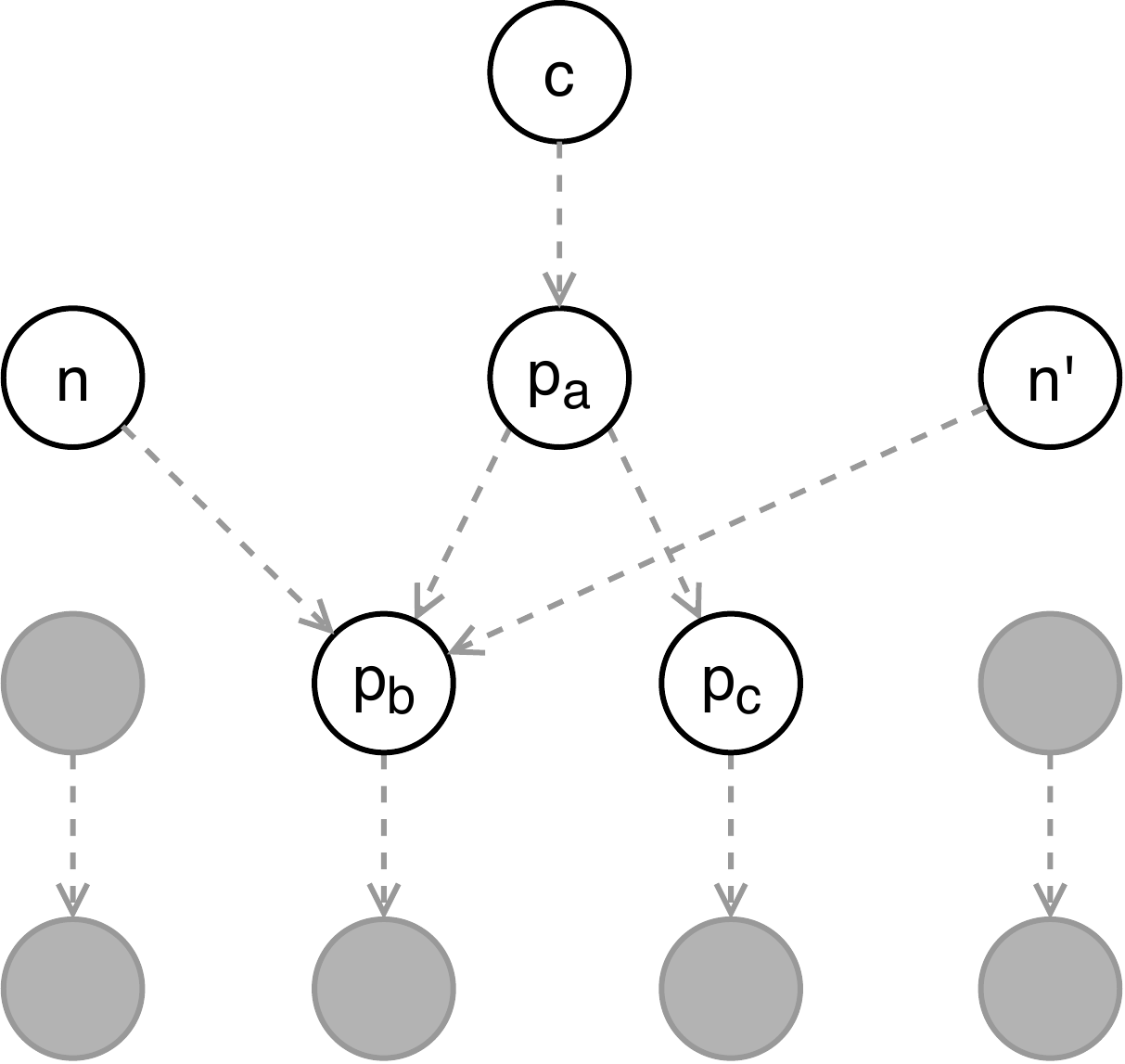}
  \caption{}
  \label{fig:overview07}
 \end{subfigure}
 \caption{An overview of system behaviour implementing our proposed solution. (a) Agent $p_b$ presents an abnormal behaviour that affects agents that depend on it. (b) Agent $c$ perceives a violation on a quality requirement and notifies agent $p_a$. (c) Agent $p_a$ replaces provider $p_b$ with ${p_b}'$ and informs $c$ that its operation is normalised. (d) $p_a$ broadcasts a request of information, which is replied by $n$ and $n'$. (e) $p_a$ notifies $p_b$ of its perceived abnormal behaviour. (f) $p_b$ informs $p_a$ that its operation is normalised. (g) $p_a$ replaces ${p_b}'$ with $p_b$ and the system returns to its former state.}
 \label{fig:overview}
\end{figure*}

We overview our interaction protocol with an example presented in Figure~\ref{fig:overview}, which illustrates a scenario where there is an abnormality caused by a service provider (an external cause).
% INGRID: component c e service c. This is confusing. melhor agent X ou services s_1, s_2, ...
In this scenario, an agent $c$ consumes a service $a$, which must comply with a given quality requirement $\mathcal{R}(q)$. This service is provided by an agent $p_a$, which consumes services $b$ and $c$ from agents $p_b$ and $p_c$, to deliver it. If there is a degradation of the quality level of the service provided by $p_b$, it is likely that this is propagated to the services provided by other agents that rely on $p_b$'s services. Therefore, agents $p_a$, $n$ and $n'$ are affected. In particular, as a consequence, $p_a$ also delivers degraded services (Figure~\ref{fig:overview01}).

Assume that $c$ detects that the service it consumes from $p_a$ violates $\mathcal{R}(q)$. This causes $c$ to send a message to $p_a$ to notify  the occurrence of such an abnormality (Figure~\ref{fig:overview02}). Receiving this notification raises on $p_a$ the need for normalising its operation. The core of our idea is that, to fulfill this need, the notified component carries out a stepwise cause diagnosis. It first verifies whether the abnormality has an internal cause. If there is no evidence of internal issues, a second verification step is performed to identify which external source is the cause of the problem. By diagnosing the source of the abnormality, $p_a$ is able to remedy and, if possible, definitely solve it, thus preventing further quality requirement violations. In our example, $p_a$ has no evidence that the abnormality comes from internal sources and, therefore, self-healing mechanisms are unable to repair it. Instead, the issue is identified as coming from the consumption of service $b$. In this case, $p_a$ remedies it by replacing the existing provider $p_b$ with ${p_b}'$ and informing $c$ that its operation was normalised (Figure~\ref{fig:overview03}), while proceeding to the second step of the verification activity.

Next, $p_a$ takes into account the data monitored by cooperating agents to determine the external cause of abnormality. To carry it out, $p_a$ broadcasts a message requesting agents that also consumed service $b$ from $p_b$ to inform, based on their history, the likelihood of receiving an anomalous reply from that provider. In our example, this request is replied by $n$ and $n'$ (Figure~\ref{fig:overview04}). After evaluating the obtained replies, $p_a$ can diagnose the problem cause and handle it accordingly. If it does not detect any abnormality coming from the suspicious agent, the communication link between them is treated as the cause, and it can act to repair it. However, if there is evidence that $p_b$ caused the abnormality, $p_a$ sends a message notifying that agent of the issue (Figure~\ref{fig:overview05}). Receiving this notification triggers on $p_b$ the same need for normalising its operation. Once it occurs, $p_a$ is informed (Figure~\ref{fig:overview06}) and becomes able to start using service $b$ from its previous provider again, thus restoring the system to its initial state (Figure~\ref{fig:overview07}). We detail the protocol that specifies how these agent interactions occur in the next section.

\subsection{Interaction Protocol}\label{sec:solution:protocol}

\begin{figure*}
 \centering
 \includegraphics[width=\linewidth]{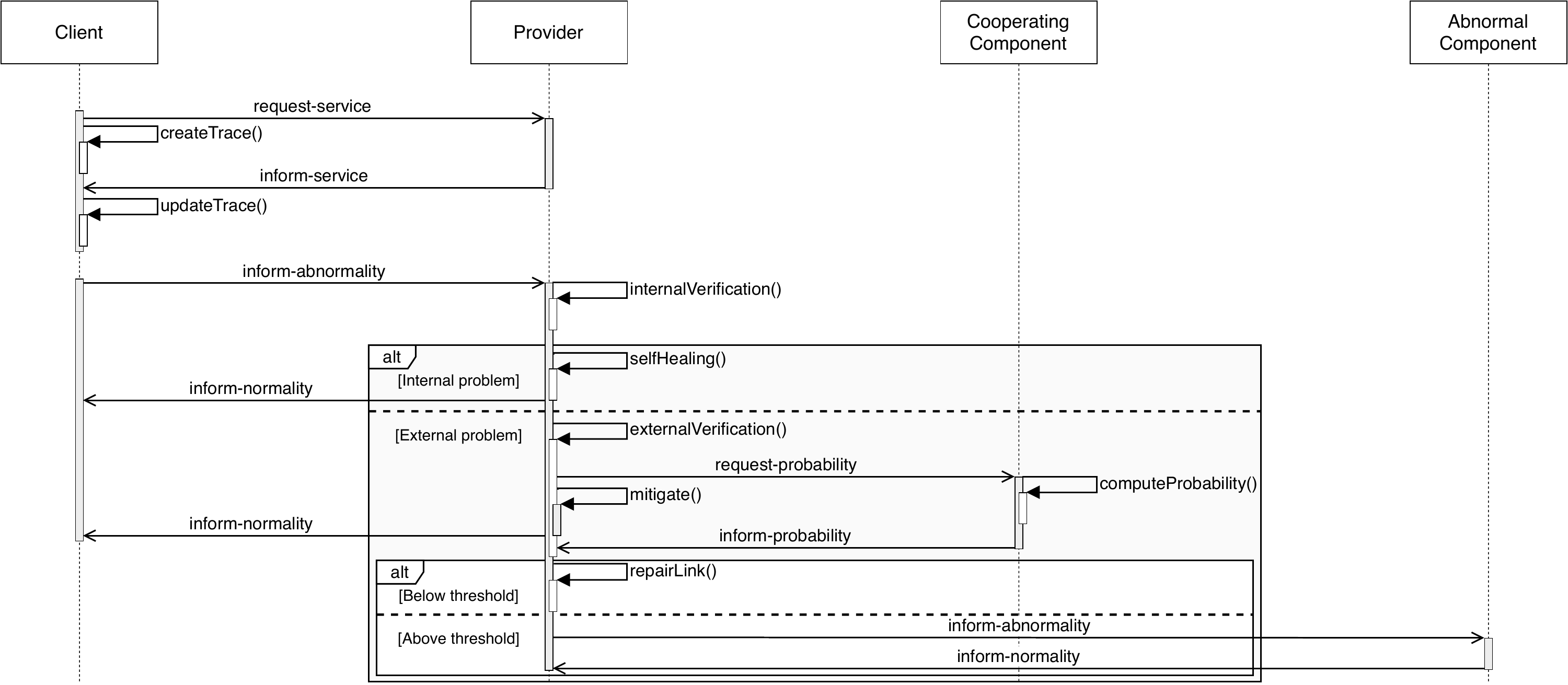}
 \caption{An interaction protocol between components of a system. Clients request services with \texttt{request-service} messages, which are replied by providers with corresponding \texttt{inform-service} messages. Clients notify abnormal components of quality requirement violations with \texttt{inform-abnormality} messages, which are replied with \texttt{inform-normality} messages after abnormalities are handled. Providers are able to issue \texttt{request-probability} messages to cooperating components, which may reply with \texttt{inform-probability} messages.}
 \label{fig:protocol}
\end{figure*}

Our protocol allows agents to exchange messages in order to collaboratively diagnose the cause of a service abnormality, assuming that agents use services provided by other agents. Figure~\ref{fig:protocol} depicts the interaction among agents. The request of a service is done by client agents, when they send a \texttt{request-service} message, which specifies the service being requested. This request can be sent to many provider agents and a negotiation may take place to choose a provider. However, this can be done with existing protocols, such as the FIPA Contract Net Protocol~\citep{fipa:02}. Our protocol thus focuses on the agent that actually provided the service. When the service has been completed, the provider replies the client request with an \texttt{inform-service} message containing the output of the corresponding service. During this exchange of messages, clients collect information that may be used later to diagnose abnormalities within the system (\texttt{createTrace()} and \texttt{updateTrace()}).

When the client detects a requirement violation on a consumed service, an \texttt{inform-abnormality} message is sent to the corresponding provider, notifying it of the quality feature and conversation in which the abnormal behaviour occurred. The provider replies this notification with an \texttt{inform-normality} message after it has taken corrective measures to address the abnormality. When this reply is sent depends on the result of the activity performed to verify whether the abnormality has an internal origin or not (\texttt{internalVerification()}). If it was caused by an internal problem, the \texttt{inform-normality} message is sent right after the execution of self-healing actions (\texttt{selfHealing()}). Otherwise, the message is sent after external causes are mitigated (\texttt{mitigate()}). This mitigation process occurs during the second step of the verification activity (\texttt{externalVerification()}), where the provider consults cooperating agents with the aim of determining which possible external cause originated the abnormality. The consultation is made through a \texttt{request-probability} message broadcast with the identification of a suspicious agent, the service under investigation, and the quality feature being considered. Recipients can reply with a refusal or an \texttt{inform-probability} message, until every agent replies or a deadline expires. The \texttt{inform-probability} message contains the likelihood of getting an anomalous reply from the suspicious agent when requesting the given service (\texttt{computeProbability()}).

From received replies, the provider computes a score that allows it to diagnose the external cause of abnormality. If the score falls below a given threshold, the communication link is considered the cause of abnormality and is handled accordingly (\texttt{repairLink()}). Otherwise, if the score is above the threshold, the provider plays the role of a client, instantiating the protocol in a new context.

\subsection{Agent Behaviour}\label{sec:solution:behaviour}

According to our protocol, an agent can act as (i) a client, by requesting services and collecting information; (ii) a provider, by delivering services and carrying out the process of verifying, remedying and solving abnormalities; and (iii) a cooperating agent, by supporting providers when required. Next, we detail these three roles.

\subsubsection{Client Agent}\label{sec:behaviour:client}

An agent plays the role of a client when it requests services from providers. During this interaction, it collects data that can also be used later when playing other roles. These data, which include information related to the performance of its providers regarding quality requirements, are collected and stored in structures named interaction traces.

\begin{definition}[Interaction Trace]
 An interaction trace $t$ is a tuple $\langle m, \mathcal{M}_q, time \rangle$, where $m$ is the traced message; $\mathcal{M}_q: \mathcal{Q} \nrightarrow \mathbb{R}$ is a partial mapping of quality features and their measured values; and $time$ is the time at which the trace was recorded.
\end{definition}

\begin{definition}[Agent]
  An agent $a$ is a tuple $\langle \mathcal{S}, \mathcal{Q}, \mathcal{R}, \mathcal{T} \rangle$, where $\mathcal{S}$ is the set of services it is able to provide; $\mathcal{Q}$ is the set of quality features it can measure; $\mathcal{R}$ corresponds to its quality requirements from the services it consumes; and $\mathcal{T}$ is the set of interaction traces collected by this component.
\end{definition}

An agent can only measure the quality of a service by using it. Therefore, only \texttt{request-service} and their corresponding \texttt{inform-service} messages are considered for tracing. After dispatching a service request, the client creates a new interaction trace through the \texttt{createTrace()} operation, and adds it to its set of traces. A recently created trace contains only the message with which it is associated. As an example, let agent $p_a$ from Figure~\ref{fig:overview} be defined as
\begin{align*}
 p_a = \langle &\{a\}, \{response\_time\}, \\
 &\mathcal{R}(response\_time) = (response\_time \leq 15), \emptyset \rangle.
\end{align*}
Let $m_0 = \langle \#1, \#1, p_a, p_b, request, b, null \rangle$, be the message sent by $p_a$ at instant 5 requesting service $b$ from $p_b$. The interaction trace $t_0 = \langle m_0, null, null \rangle$ is thus created and added to $p_a$'s set of traces, resulting in 
\begin{align*}
 p_a = \langle &\{a\}, \{response\_time\}, \\
 &\mathcal{R}(response\_time) = (response\_time \leq 15), \{t_0\} \rangle.
\end{align*}
We assume every agent has the mechanisms required to measure the performance of providers regarding its quality requirements. Therefore, after receiving a reply to a service request, the client updates the previously created trace with the measured provider's performance as well as the time at which the reply was received and, consequently, the service was provided. It is done through the \texttt{updateTrace()} operation. In our example, $p_a$ has a single quality requirement that states that consumed services must be delivered ($response\_time$) within 15 units of time. After receiving the reply from $p_b$ at instant 12, $p_a$ updates its trace $t_0$ to $t_0 = \langle m_0, \mathcal{M}_q(response\_time) = 7, 12 \rangle$, indicating that its request was fulfilled in 7 units of time. Constantly tracing their interactions allows clients to build a dataset that represents the performance of their providers over time.

Clients also notify service providers of quality requirement violations. These notifications can be either the result of a self-awareness mechanism or a reaction to a notification received by the client when playing the role of provider. While the former is not the focus of this work, the latter is detailed next, as we describe the behaviour of provider agents.

\subsubsection{Provider Agent}\label{sec:behaviour:provider}

An agent acting as a provider has the main goal of delivering services to clients with the expected quality. It is done by replying to service requests with the corresponding service output. However, when this goal is not achieved and the provider is notified that a delivered service was provided with an abnormality, a new goal is generated in order to normalise its operation.

In order to achieve this goal, the provider puts into action a strategy to diagnose, remedy and solve the cause of the abnormality. As mentioned in Section~\ref{sec:solution:protocol}, the first step of this strategy comprises the \texttt{internalVerification()} activity, which is responsible for checking if the abnormality was introduced by the provider itself. The problem is considered to have an internal cause if, to deliver its service, the provider did not consume any service supplied with a largely different performance than what was usually recorded on its set of interaction traces. Algorithm~\ref{alg:internal-verification} describes how this activity is carried out. Auxiliary functions are presented in Table~\ref{tbl:auxFunctions}.

\begin{algorithm}[t]
\caption{Internal verification activity.}\label{alg:internal-verification}
\begin{algorithmic}[1]
\Require the set $T$ of interaction traces, the violated quality requirement feature $q$, the identifier $id_c$ of the conversation in which the violation was perceived.
\Ensure a message informing that the agent operation is normalised.
\item[]
\small
\Procedure{internalVerification}{$T, q, id_c$}
    \State $I_a\gets \emptyset$
    \State $T'\gets \Calls{getTraces}{T, id_c}$
    \For{$t \in T'$}
        \State $s\gets t[m][s]$
        \State $p\gets t[m][c_r]$
        \State $time\gets t[time]$
        \State $L_{\mathcal{M}_q}\gets \Calls{getMeasurements}{T, s, p, q, time}$
        \If{$\Calls{isAnomalous}{L_{\mathcal{M}_q}}$}
            \State $id_m\gets t[m][id_m]$
            \State $I_a\gets I_a \cup \langle s, p, id_m \rangle$
        \EndIf
    \EndFor 
        \If{$I_a = \emptyset$} \Comment{Internal cause}
            \State \Calls{selfHealing}{}
            \State $\Broadcast{inform-normality}{}$
        \Else \Comment{External cause}
            \For{$\langle s, p, id_m \rangle \in I_a$}
                \State $\Calls{mitigate}{s}$
                \State $\Broadcast{inform-normality}{}$
                \State $score\gets \Calls{externalVerification}{s, p, q}$
                \If{$score \leq threshold$} \Comment{Link issue}
                    \State \Calls{repairLink}{}
                \Else \Comment{Provider issue}
                    \State $\Send{inform-abnormality}{q, id_m}$
                    \State $\Receive{inform-normality}$
                \EndIf
                \State $\Calls{undo}{}$
            \EndFor
        \EndIf
\EndProcedure
\end{algorithmic}
\end{algorithm}

\begin{table*}
	\caption{Description of Auxiliary Functions}\label{tbl:auxFunctions}
	\centering
	\small
	 \renewcommand\tabularxcolumn[1]{m{#1}}
	 \begin{tabularx}{\textwidth}{l>{\centering}p{3cm}X}
		\hline
		\textbf{Function} & \textbf{Mathematical Expression} & \textbf{Description} \\ \hline
		\texttt{getTraces($T, id_c$)} & $\begin{aligned}
			\{ t \colon t &\in T \wedge \\ m &= t[m] \wedge \\ id_c &= m[id_c]\}
                \end{aligned}$ & Gives the set of traces recorded during conversation with identification $id_c$.
		\\ \hline
		\texttt{getMeasurements($T, s, p, q, time$)} & $\begin{aligned} 
            \{ \mathcal{M}_q(q) \colon t &\in T \wedge \\ 
            m &= t[m] \wedge \\ 
            p &= m[c_r] \wedge \\ 
            s &= m[s] \wedge \\  
            \mathcal{M}_q &= t[\mathcal{M}_q] \wedge \\ 
            time &\geq t[time]\} 
                \end{aligned}$ & Gives the list of measurements of quality feature $q$ obtained from service $s$ delivered by provider $p$ before time $time$.
		\\ \hline
		\texttt{getTimes($T, s, p, time$)} & $\begin{aligned} 
            \{ t[time] \colon t &\in T \wedge \\ 
            m &= t[m] \wedge \\  
            p &= m[c_r] \wedge \\ 
            s &= m[s] \wedge \\ 
            time &\geq t[time]\} 
                \end{aligned}$ & Gives the list of times at which service $s$ was delivered by provider $p$ before time $time$.
		\\ \hline
	 \end{tabularx}
\end{table*}%

Initially, the \texttt{getTraces($T, id_c$)} function retrieves the traces recorded during the abnormal conversation $id_c$ (line 3). Service $s$ and its provider $p$ are identified for each trace, and the list of all measurements of quality feature $q$ that were taken when consuming $s$ from $p$ until the time the abnormal conversation occurred is retrieved (lines 5--8)---see \texttt{getMeasurements($T, s, p, q, time$)} function in Table~\ref{tbl:auxFunctions}.

A statistical analysis of the resulting list of measurements is performed in order to check whether $p$'s performance on the abnormal conversation $id_c$ can be classified as anomalous when compared to historical data. This analysis, executed within the \texttt{isAnomalous($L_{\mathcal{M}_q}$)} function (line 9), applies a method called \emph{Tukey's fences}~\citep{tukey:77} to compute the lower and upper boundaries---the fences---that determine the range of normal values from a sample. Any value laying outside these boundaries is considered an outlier and, consequently, an anomaly. To compute these fences, this method takes as input a list of values arranged in an ascending order and identifies its lower ($Q_1$) and upper ($Q_3$) quartiles. $Q_1$ corresponds to the median of the range of values below the median of the entire sample, while $Q_3$ stands for the median of the range of values above the median of the entire sample. Eq.~\ref{eq:lowerBound} and \ref{eq:upperBound} are then applied to calculate the lower ($b_l$) and upper ($b_u$) boundaries, respectively.

\begin{align}
 b_l = Q_1 - 1.5 * (Q_3 - Q_1) \label{eq:lowerBound} \\
 b_u = Q_3 + 1.5 * (Q_3 - Q_1) \label{eq:upperBound}
\end{align}

The \texttt{isAnomalous($L_{\mathcal{M}_q}$)} function thus returns a boolean value that indicates if the last measurement from $L_{\mathcal{M}_q}$ is lower than $b_l$ or greater than $b_u$ when the \emph{Tukey's fences} method takes $L^s_{\mathcal{M}_q}$ as input, being $L^s_{\mathcal{M}_q}$ the result of sorting $L_{\mathcal{M}_q}$ in ascending order. As an example, let a list of quality feature measurements $L_{\mathcal{M}_q}$ be $L_{\mathcal{M}_q} = (8, 7, 11, 8, 8, 9, 47)$. In this list, $Q_1$ is equal to 8, and $Q_3$ is equal to 10. Therefore, according to Eq.~\ref{eq:lowerBound} and~\ref{eq:upperBound}, $b_l = 5$ and $b_u = 13$. As a result, \texttt{isAnomalous($L_{\mathcal{M}_q}$)} would return true, as 47 is greater than the obtained upper boundary.

The interaction comprising the service $s$, its provider $p$, and the message identifier $id_m$ of each trace whose measurement is classified as anomalous is added to a set of anomalous interactions (lines 9--11). If by the end of this first verification step no anomalous interaction is identified, the abnormality cause is considered to be internal. The abstract \texttt{self-healing()} operation is thus performed to repair the issue, normalising the agent behaviour, which is broadcast to the system (lines 12--14). Otherwise, if any interaction is classified as anomalous, the cause is considered external to the provider. In this case, for each anomalous interaction, an abstract \texttt{mitigate($s$)} operation is carried out. This activity is expected to remedy the external issue, thus normalising the agent behaviour and similarly broadcasting this to the system (lines 15--18). \texttt{self-healing()} and \texttt{mitigate($s$)} are abstract operations because they have the implementation that is suitable to satisfy the needs of the target domain.

Having the external cause mitigated, the second verification step takes place to identify which of the external sources, namely, the abnormal provider $p$ or the communication link, is the current cause of the issue. This verification is carried out by the \texttt{externalVerification($s, p, q$)} operation, whose outcome is a score for the performance of $p$ when providing service $s$ with respect to quality feature $q$ (line 19). If the obtained $score$ is less than or equal to a predefined $threshold$ (in this work, we use the value of 0.5, or 50\%), the communication link is considered the source of abnormality, and is thus repaired by the abstract \texttt{repairLink()} operation (lines 20--21). If the score is greater than the threshold, the provider $p$ is considered the problem cause. This conclusion follows the rationale that an unusual behaviour from a suspicious provider may be perceived not only by the agent handling the abnormality, but also by cooperating agents that consumed the same service. The provider handling the issue acts as a client, notifying $p$ of its abnormal performance regarding quality feature $q$ when replying to message $id_m$, and waiting that agent to inform when its operation gets normalised (lines 22--24). Finally, when the problem cause is solved, the abstract \texttt{undo()} operation is performed to revoke any reversible measure taken to remedy the problem cause (line 25). An existing strategy~\citep{faccin:18b} can be used as a concrete implementation of this operation.

\begin{algorithm}
\caption{External verification activity.}\label{alg:external-verification}
\begin{algorithmic}[1]
\Require the abnormal service $s$, the suspicious provider $p$, the violated quality requirement feature $q$.
\Ensure a score $score$ for the performance of $p$ be abnormal regarding quality feature $q$ when delivering service $s$.
\item[]
\Procedure{externalVerification}{$s, p, q$}
    \State $Inf\gets \emptyset$
    \State \Broadcast{request-probability}{$s, p, q$}
    \While{$\neg condition$}\Comment{Deadline or \# of replies}
        \State $m\gets \Receive{inform-probability}$
        \State $c_s\gets m[c_s]$
        \State $prob\gets m[cont]$
        \State $Inf\gets Inf \cup \langle c_s, prob\rangle$
    \EndWhile
    \State $prob_{w}\gets 0.0$
    \State $idx_s\gets 0.0$
    \For{$\langle c_s, prob\rangle \in Inf$}
        \State $prob_{w}\gets prob_{w} + (prob \times \mathcal{I}(this, c_s))$
        \State $idx_s\gets idx_s + \mathcal{I}(this, c_s)$
    \EndFor
    \State $score\gets prob_{w} / idx_s$
    \State \Return{score}
\EndProcedure
\end{algorithmic}
\end{algorithm}

First, a message is broadcast to the system requesting to cooperating agents the probability of provider $p$ to present an abnormal behaviour when delivering service $s$ with respect to quality feature $q$ (line 3). Replies are received until a stop condition is met, which can be either a deadline or a predefined number of replies. For each received reply, its sender $c_s$ and the informed probability $prob$ are added to a set $Inf$ containing received information (lines 4--8).

Agents may differ from each other regarding several characteristics. Consequently, more importance is given to information from more similar agents. This is taken into account to combine received probabilities into a final score. The similarity index $\mathcal{I}: \mathcal{C} \times \mathcal{C} \rightarrow [0, 1]$ is thus a function that maps a pair of agents $p, q \in \mathcal{C}$ to a value indicating the similarity between $p$ and $q$. 0 and 1 are the lower and highest similarities, respectively. In this work, this index is inversely proportional to the distance, measured in hops, between two agents. We assume that every agent has access to this information. The score of a suspicious provider $p$ is computed as the average of probabilities given by cooperating agents weighted by their similarity indexes (lines 9--14), and returned at the end of this activity (line 15).

\subsubsection{Cooperating Agent Behaviour}\label{sec:behaviour:cooperating}

A cooperating agent aims to provide required information to agents handling an abnormal behaviour. This is achieved by the computation of the probability of a suspicious agent to provide an anomalous measurement of a given quality feature when delivering a particular service. Algorithm~\ref{alg:probability} describes how this activity is performed. 

\begin{algorithm}[ht]
\caption{Probability computation activity.}\label{alg:probability}
\begin{algorithmic}[1]
\Require the service $s$, the provider $p$, the quality requirement feature $q$.
\Ensure a probability $prob$ of $p$ to provide an anomalous measurement of $q$ when delivering service $s$.
\item[]
\Procedure{computeProbability}{$s, p, q$}
    \State $L_{\mathcal{M}_q}\gets \Calls{getMeasurements}{T, s, p, q, now}$
    \State $L_{time}\gets \Calls{getTimes}{T, s, p, now}$
    \If{$L_{\mathcal{M}_q} \not= \emptyset$}
        \State $f\gets \Calls{getFunction}{L_{\mathcal{M}_q}, L_{time}}$
        \State $prob\gets \Calls{integrate}{f, L_{\mathcal{M}_q}}$
        \State \Send{inform-probability}{$prob$}
    \EndIf
\EndProcedure
\end{algorithmic}
\end{algorithm}

First (lines 2--3), the lists of (i) all measurements of a quality feature $q$ taken when consuming service $s$ from $p$, and (ii) the corresponding times in which these measurements were recorded, are retrieved by the functions \texttt{getMeasurements($T, s, p, q, time$)} and \texttt{getTimes($T, s, p, time$)}---see Table~\ref{tbl:auxFunctions}. If, at any time, this agent consumed $s$ from $p$, the probability computation is carried on with the identification of the probability density function $f$ that describes the recorded quality feature measurements. In our approach, given that $f$ is unknown and available measurements constitute a sample from it, $f$ is estimated through the use of a kernel density estimator.

A \emph{kernel density estimator} (KDE)~\citep{parzen:62} is a non-parametric statistical method able to approximate $f$ using a mixture of kernels $K$, each of them centred at the points $x_i$ of an available dataset---in our case, the list $L_{\mathcal{M}_q}$ of quality feature measurements. Typically used kernels include Gaussian and Epanechnikov, although any symmetric probability density function can be adopted. A bandwidth $h$ is used to acknowledge the existence of an unknown density of points in the neighbourhood of each point $x_i$, while weights $w_i$ are used to consider that a point $x_i$ may have higher surrounding densities than other points. The resulting probability density function $\hat{f_{h}}$ can thus be used to estimate the probability of any point $x$. A KDE is thus defined as
\begin{equation}
 \hat{f_{h}}(x) = \sum_{i=1}^n w_i K_h(x - x_i),
\end{equation}
where $K_h(x) = K(\nicefrac{x}{h}) / h$ for kernel $K$ and bandwidth $h$, and $\sum_{i=1}^{n} w_i = 1$.
Recently collected measurements may have higher informative power than those earlier collected. Therefore, the weight $w_i$ assigned to each measurement $x_i \in L_{\mathcal{M}_q}$ is proportional to the recency of that measurement. This is calculated according to Eq.~\ref{eq:weight}. 

\begin{equation}\label{eq:weight}
 w_i = L_{time_i} \times \frac{1}{\sum_{j=1}^{|L_{time}|} L_{time_j}}
\end{equation}
The obtained function $\hat{f_{h}}$ thus comprises the result of executing the \texttt{getFunction($L_{\mathcal{M}_q}, L_{time}$)} operation (line 5). 

Finally, the \texttt{integrate($f, L_{\mathcal{M}_q}$)} operation is executed to determine the probability $prob$ of obtaining an anomalous measurement from $p$ (line 6). In this operation, the \emph{Tukey's fences} method (Section~\ref{sec:behaviour:provider}) is applied to the list $L_{\mathcal{M}_q}$ of measurements with the aim of identifying the lower ($b_l$) and upper ($b_u$) boundaries that delimit the range of normal values. One minus $f$  integrated using these boundaries (Eq.~\ref{eq:integrate}) gives the estimated distribution. It is the probability $prob$ of a random value from that distribution to fall in a range below $b_l$ or above $b_u$, thus becoming an anomalous value. At the end, this obtained probability is informed to the agent that requested it (line 7).

\begin{equation}\label{eq:integrate}
 prob = 1 - \int_{b_l}^{b_u} f
\end{equation}

\begin{figure}[t]
 \centering
 \includegraphics[width=0.6\linewidth]{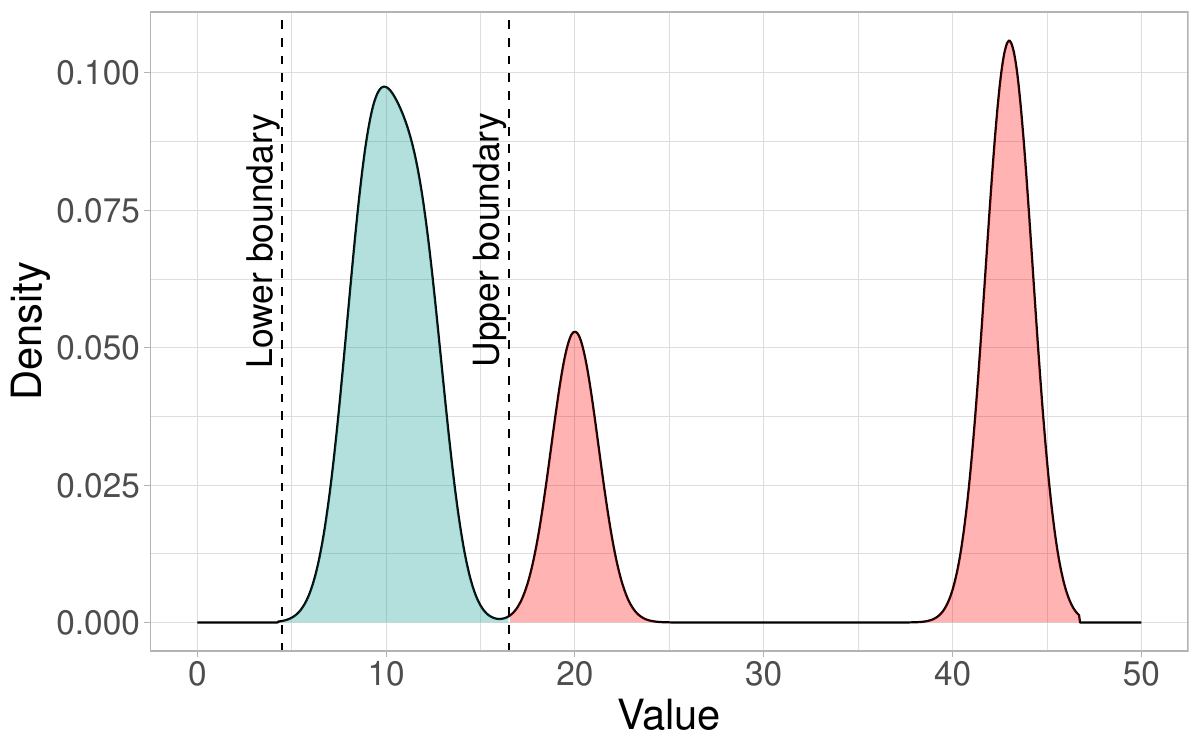}
 \caption{The estimated probability distribution obtained after executing \texttt{getFunction($L_{\mathcal{M}_q(response\_time)}, L_{time}$)}. Values laying outside the lower and upper boundaries are considered anomalous.}
 \label{fig:exampleKDE}
\end{figure}

To exemplify this process, let $n$ be a cooperating agent that receives a request to provide the probability of $p_b$ behaving abnormally when delivering service $b$ with respect to $response\_time$. Let 
\begin{equation*}
L_{\mathcal{M}_q(response\_time)} = (8, 10, 9, 9, 11, 12, 10, 9, 12, 20, 43)
\end{equation*} 
be the list of quality feature measurements and
\begin{equation*}
L_{time} = (5, 10, 15, 20, 25, 30, 35, 40, 45, 50, 55)
\end{equation*} the list of times at which these measurements were registered. The estimated probability distribution $f$ obtained by $n$ after performing the \texttt{getFunction($L_{\mathcal{M}_q(response\_time)}, L_{time}$)} operation is depicted in Figure~\ref{fig:exampleKDE}. In our example, the lower and upper boundaries of $L_{\mathcal{M}_q(response\_time)}$ are equal to 4.5 and 16.5, respectively. It means that, while any value that falls within these boundaries are considered normal (blue shaded in Figure~\ref{fig:exampleKDE}), values that fall in any region outside them comprise an anomaly (red shaded area). After carrying out the \texttt{integrate($f, L_{\mathcal{M}_q(response\_time)}$)} operation, $n$ verifies that the probability $prob$ of obtaining an anomalous measurement when requesting service $b$ from $p_b$ with respect to $response\_time$ is approximately 0.49 or 49\%.

% !TEX root = main.tex

\section{Evaluation}\label{sec:evaluation}

Having detailed our protocol and the behaviour of each role, we now evaluate our proposal with an empirical evaluation. In this evaluation we examine the performance of a system that implements our cooperative strategy regarding its ability to diagnose the root cause of problems in different locations and coming from different sources. We compare the obtained results with the performance of the same system when applying two other resilience strategies. In the first one, which we named \emph{passive strategy}, agents ignore any notification of abnormality sent to them. In the second one, named \emph{remedial strategy}, these agents are able to mitigate abnormalities but do not have the mechanisms to diagnose and solve their causes. Even though there are many existing solutions focused on the diagnosis and resolution of problems causes in distributed systems, they all differ from our approach in some key aspects. First, most of them do not address both facets of the problem, focusing either only on the diagnosis or on the problem solving aspect. Solutions capable of diagnosing problem causes differ from ours mainly regarding the size and complexity of the scenarios they are able to handle, being inefficient when dealing with scenarios comprising more than a fixed (and considerably low) number of agents. Solutions focused on problem solving, in turn, usually rely on centralised information in order to deal with challenging situations. Considering these characteristics, a direct comparison with existing solutions is not meaningfull. Next, we present the procedure adopted in our evaluation, followed by its results and discussion.

\subsection{Procedure}

\begin{figure*}[t]
 \centering
 \includegraphics[width=\linewidth]{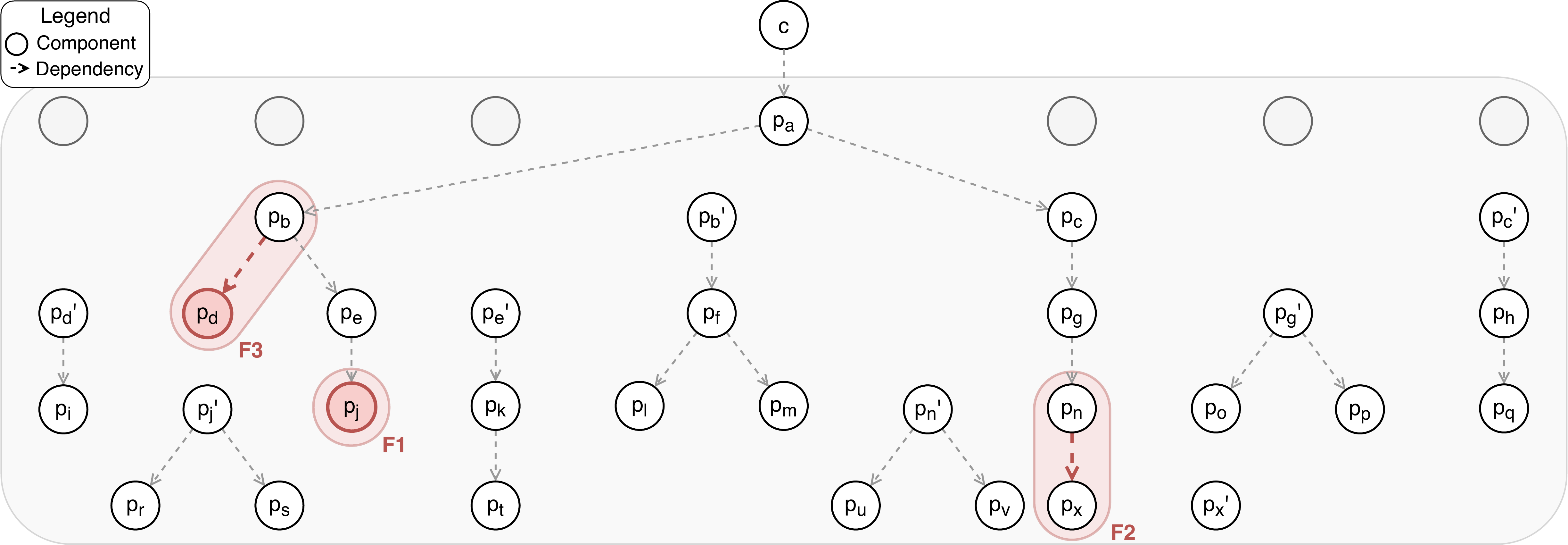}
 \caption{A service-oriented system comprising 37 autonomous components and an external client $c$. \emph{F1}, \emph{F2} and \emph{F3} are failures introduced to the system to affect different components and communication links.}
 \label{fig:scenario}
\end{figure*}

Many MAS are open, so agents can join or leave the system at runtime, leading to dynamic scenarios. Consequently, there are no static agent topologies used as a standard for simulations. It is thus a common practice to elaborate synthetic topologies and workloads to simulate realistic scenarios in order to evaluate the proposed approaches~\citep{Dotterl:17}. To evaluate our approach, we implemented the scenario depicted in Figure~\ref{fig:scenario}. It represents a service-oriented system comprising 37 autonomous components able to interact with each other by consuming and delivering services, and one external component acting as a top-level client. Services of different providers have varying costs, which means that the same service may be more or less expensive when requested from different sources. With the exception of unnamed components, which use services from any provider despite of their costs, every other component has a secondary goal of minimising the cost associated with consuming services. Dependencies represented on Figure~\ref{fig:scenario} indicate the initial providers used by each agent.

As introduced, our approach includes abstract activities that are specific to concrete implementations. In this experiment, the \texttt{selfHealing()} and \texttt{repairLink()} activities are simulated. In real world scenarios, these operations are expected to comprise, e.g., the implementation of one of the many existing self-healing, self-configuration or repairing strategies~\citep{ghosh:07}. The implemented \texttt{mitigate($s$)} operation, in turn, replaces the current (abnormal) provider of the given service $s$ with a different (and presumably more costly) one. An alternative implementation of this operation may be, for instance, a prediction of the output of a given service based on the current context. Finally, the stop condition for providers to wait for replies from cooperating agents in the \texttt{externalVerification($s, p, q$)} operation is set to a 5 seconds deadline.

The external agent $c$ has a quality requirement $\mathcal{M}_q(response\_time) = (response\_time \leq 250)$, which specifies that the services it consumes must be provided in no longer than 250 ms. In our experiment, services have an initial delivery time of 10 ms. This time increases by 250 ms for each failure affecting the service provision, which can occur either on the service provider or in the communication link between it and its client.

Our experiment thus consists of measuring the cost and response time observed by $c$ when consuming service $a$ from $p_a$. We ran a simulation with 120 episodes, each of them comprising the following sequence of steps: (i) $c$ requests service $a$ from $p_a$; (ii) unnamed components request different services within the system; and (iii) the cost and response time measured by $c$ are recorded. Three failures, \emph{F1}, \emph{F2}, and \emph{F3}, are injected on the simulation at the 30th, 60th and 90th episode, respectively, with the aim of causing abnormalities on service provision. In \emph{F1}, agent $p_j$ is targeted, increasing its response time and becoming unable to provide its service as usual. F2 affects the link between agents $p_n$ and $p_x$, thus slowing their communication down. Finally, F3 disturbs both the functioning of agent $p_d$ and its link with $p_b$. These points of failure are depicted in Figure~\ref{fig:scenario}.

\subsection{Results and Discussion}
\begin{figure}[t]
    \centering
    \begin{subfigure}{0.45\textwidth}
        \includegraphics[width=\textwidth]{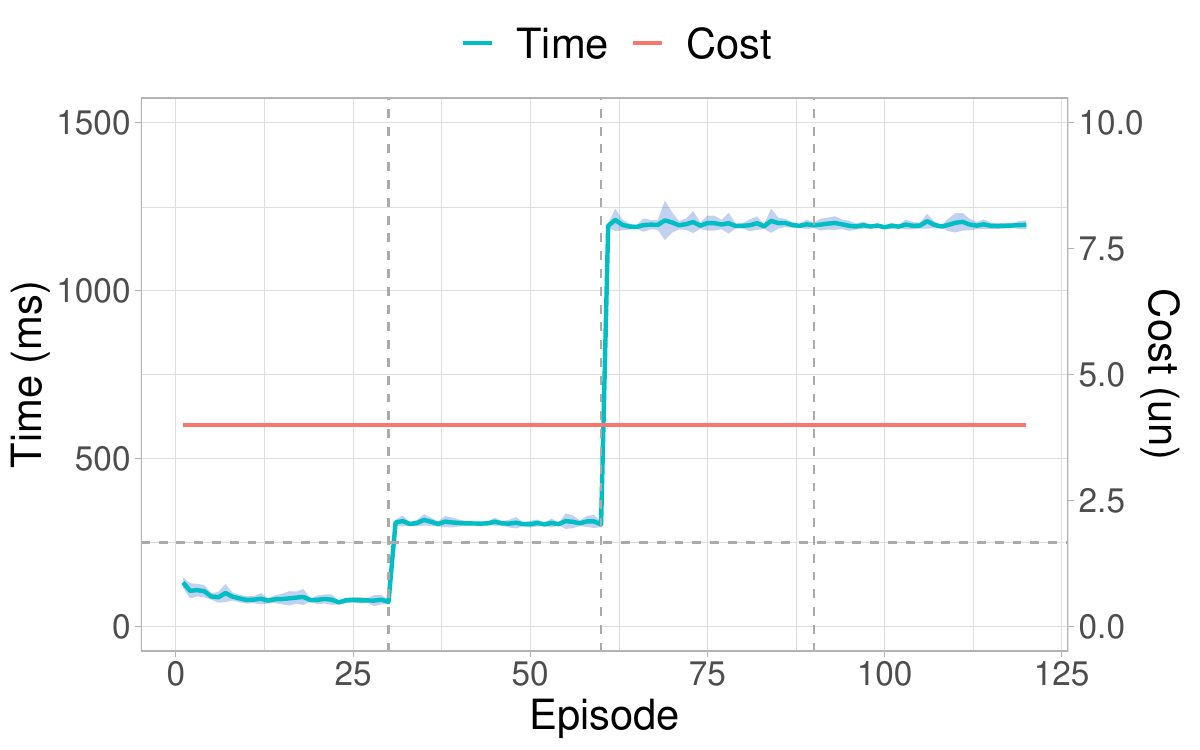}
        \caption{Passive strategy}
        \label{fig:results:passive}
    \end{subfigure}
    \begin{subfigure}{0.45\textwidth}
        \includegraphics[width=\textwidth]{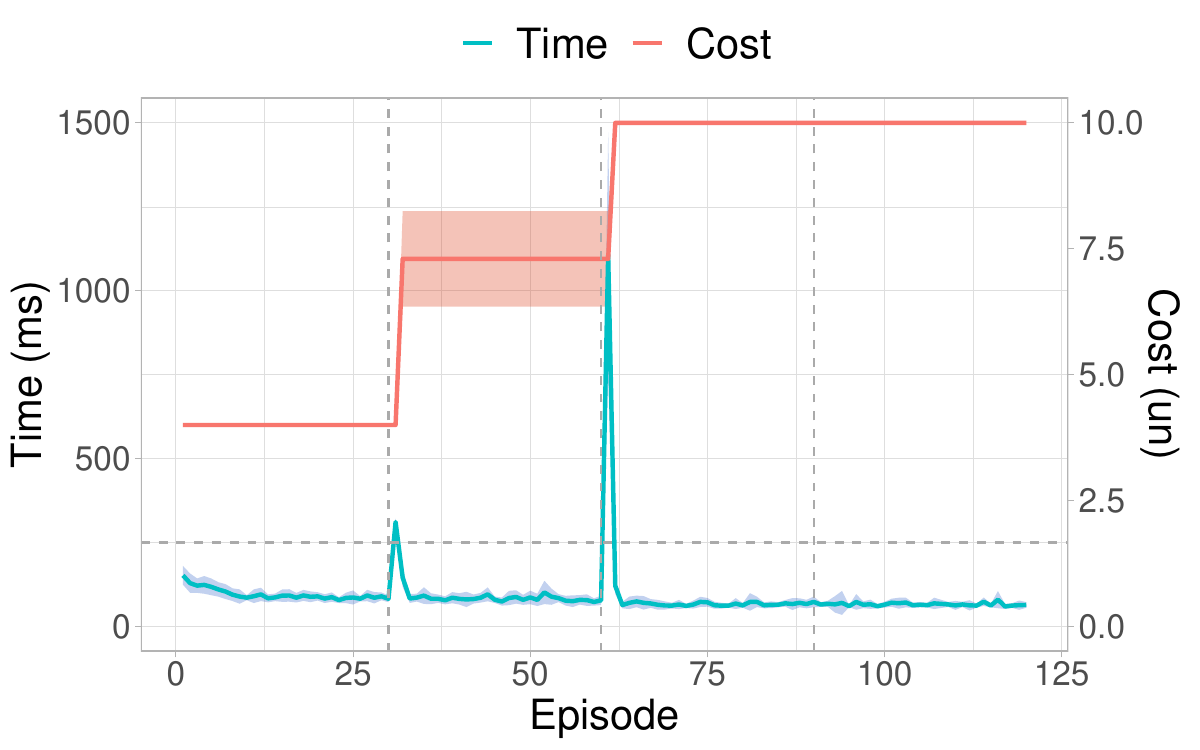}
        \caption{Remedial strategy}
        \label{fig:results:remedial}
    \end{subfigure}\\
    \begin{subfigure}{0.45\textwidth}
        \includegraphics[width=\textwidth]{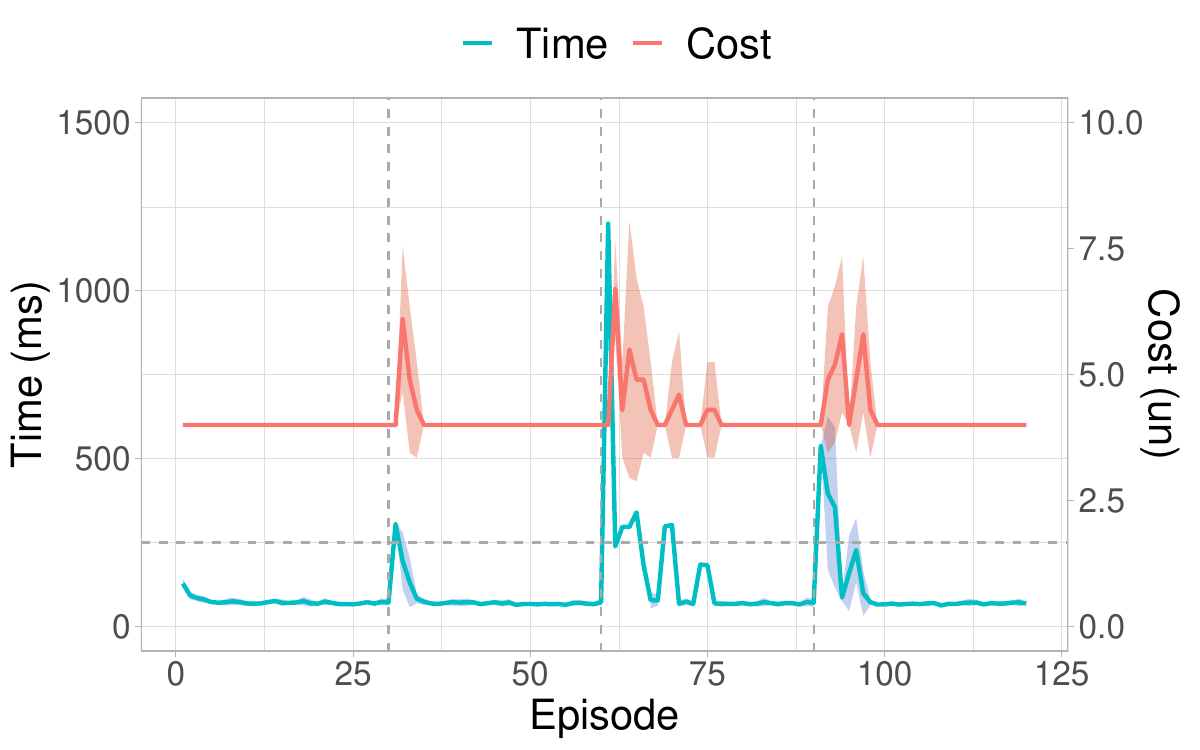}
        \caption{Cooperative strategy}
        \label{fig:results:cooperative}
    \end{subfigure}
    \caption{Simulation results (time and costs by episode).}\label{fig:results}
\end{figure}

We executed our simulation 10 times for each adopted problem-solving strategy. The results are discussed focusing on the system behaviour observed during the execution of our experiment. These results are presented in Figure~\ref{fig:results}, where blue and red lines represent the average response time and average cost observed by agent $c$ in each episode, respectively, and shaded areas represent their corresponding standard deviation.

Figure~\ref{fig:results:passive} presents the results obtained by adopting the passive strategy. It is possible to notice that, even with slightly higher values caused by the initialisation procedure at the beginning of the simulation, the response time measured by $c$ maintains an average of $\approx$85 ms until the occurrence of the first failure. It is important to highlight that, as in our implementation agents have a single execution thread---meaning that service requests are queued in order to be fulfilled---the response time perceived by a client may vary according to the amount of requests being processed by its provider. This explains why the measurement recorded by $c$ when consuming $a$ from $p_a$ is usually greater than the 40 ms that would be expected if there were no other agents consuming that and other related services.

After the introduction of \emph{F1}, the response time increases and stays above the quality requirement threshold at an average of 308.5 ms. It increases again after the occurrence of \emph{F2} and remains stable from that moment, even with the occurrence of the third failure, which ends up being subsumed by the former two. Such an outcome takes place because, when provider $p_a$ adopts the passive strategy, it simply ignores notifications of abnormality received from $c$. As a consequence, in order to deliver $a$, that agent keeps consuming services from providers affected by the introduced failures. The observed cost also reflects such a behaviour, standing steady for the entire simulation due to the fact that initial providers are not replaced with others.

The scenario changes completely when we consider results from simulations in which the remedial strategy is adopted. They are shown in Figure~\ref{fig:results:remedial}. After the occurrence of \emph{F1}, $p_a$'s response time increases and surpasses the quality threshold, peaking at an average of 311 ms, but quickly decreasing and stabilising at an average of $\approx$85 ms. It happens because, by following the remedial strategy, agents are able to mitigate abnormalities coming from external sources by changing their service providers. Consequently, after receiving a notification of abnormality from $c$, $p_a$ replaces $p_b$ with ${p_b}'$, an agent that does not make use---neither directly or indirectly---of the abnormal service provided by $p_j$ in order to deliver $b$. It is also because of this provider replacement that the cost increases. The introduction of \emph{F2} results on the same behaviour pattern, with response time increasing above the quality threshold, peaking at 740.6 ms, and decreasing after provider $p_c$ is replaced with ${p_c}'$, which also results in a higher associated cost. It is interesting to notice that the occurrence of \emph{F3} does not affect the performance of $p_a$ perceived by $c$. It is explained by that fact that $p_a$ is no longer a client of $p_b$ and, as a consequence, degraded services provided by that agent are not consumed.

Considering that the goals of agent $c$ are to consume service $a$ while satisfying its quality requirements and minimising the associated cost, our cooperative strategy presents an optimal trade-off towards the achievement of both objectives. The outcomes with this strategy are shown in Figure~\ref{fig:results:cooperative}. After \emph{F1}, it is possible to observe the increasing response time, which peaks at 305 ms and decreases after that, following the same pattern presented when adopting the remedial strategy. In contrast to this strategy, the measured cost returns to its initial level after an increase due to the adopted mitigation. Such a behaviour is explained by the propagating nature of our approach as well as the existence of the \texttt{undo()} operation in our strategy.

When the first failure affects $p_j$ and its degraded performance spreads to its clients, $p_a$ is notified of an abnormality. It thus carries out the cause verification operation and identifies the external nature of the issue, which is remedied with the replacement of $p_b$ with ${p_b}'$. Because of this change, the response time perceived by $c$ returns to an acceptable level, but at the expense of an increased cost. After consulting the cooperating---unnamed---agents, and identifying the external source of the issue, $p_a$ notifies $p_b$ of the detected abnormality. That agent thus carries out the same process, diagnosing the external origin of the problem, mitigating it and notifying the suspicious provider of its abnormality. After mitigating the problem cause, $p_b$ informs $p_a$ that its operation was normalised, which allows $p_a$ to revert the remediation action taken and go back to consuming service $b$ from $p_b$, thus reducing its cost. Meanwhile, $p_e$ and $p_j$ performs the same identification process after being notified of their abnormalities by $p_b$ and $p_e$, respectively. When $p_j$ diagnoses that its degraded performance was caused by an internal failure, the \texttt{selfHealing()} operation is executed, normalising the agent behaviour and allowing the system to return to its initial setup.

The same process is executed by the system when failures \emph{F1} and \emph{F2} are introduced. In Figure~\ref{fig:results:cooperative}, it is possible to notice some peaks after the occurrence of each of these failures. It happens because, in some executions, cooperating components are unable to immediately classify some suspicious agents as abnormal ones, which results on the incorrect problem cause being addressed. For instance, consider that, after the occurrence of \emph{F2}, agent $p_c$ mitigated an external cause by replacing $p_g$ with ${p_g}'$ and, by consulting its cooperating agents, it (mistakenly) decided that $p_g$ did not present any abnormality. In this case, $p_c$ would repair its communication link with $p_g$, acknowledge the problem as solved, and go back to consume services from that agent. As a consequence, the abnormality would reappear and be notified once again.

Nevertheless, this situation demonstrates that the execution of some additional episodes is sufficient to allow cooperating agents to correctly recognise the existence of abnormalities on suspicious components and to provide conditions for the system to find the correct root cause of its problems. Therefore, even with the introduction of three failures from different sources at different levels, at the end of our simulation the system is able to correctly diagnose and solve failure causes, with agent $c$ satisfying its quality requirements at a reduced cost.

Finally, as shown in Table~\ref{tbl:results:summary}, simulations in which the system adopts our cooperative strategy took, on average, 616.5 seconds to be executed, which represents a reduction of 10.57\% and 0.32\% when compared to the adoption of passive (689.4 s) and remedial (618.5 s) strategies, respectively. The accumulated cost presented by our approach, in turn, was 3.81\% higher (498.3 unities of cost) when compared to the passive strategy (480.0 unities of cost), and 46.59\% lower in comparison to the remedial strategy (933.0 unities of cost).

\begin{table}
 \centering
 \caption{Average Execution Time and Accumulated Cost}\label{tbl:results:summary}
 \begin{tabular}{lcc}
  \toprule
  \textbf{Strategy} & \textbf{Execution Time (s)} & \textbf{Accumulated Cost (un)} \\
  \midrule
  Passive  & 689.4 & 480.0 \\ 
  Remedial & 618.5 & 933.0 \\ 
  Cooperative & 616.5 & 498.3 \\ 
  \bottomrule
 \end{tabular}
\end{table}

These results show that our approach is able to combine the benefits from both, passive and remedial strategies, into a single solution. However, to achieve this outcome there are some prerequisites that must be considered. First, as already mentioned, it is required that cooperating agents have consumed services from a suspicious agent after the occurrence of an abnormality in order to be able to correctly diagnose it. Otherwise, they could provide information that does not represent the current behaviour of that provider and, consequently, the agent handling the abnormality would not be able to correctly diagnose the source of the issue. Strategies to overcome this limitation may include, for instance, enforcing cooperating agents to issue a service request in order to update their sets of interaction traces, or limit agents able to cooperate to those whose last requirements to a suspicious provider occurred within a given time interval. The applicability of these strategies and their impact on our solution are subject to further investigation. 

Our approach also relies on the quality of the remedial actions taken by the agent during the process of mitigating external causes. In our evaluation, we replace service providers to normalise system operation while the root cause of the problem is identified and solved. If, for instance, the temporary provider also exhibits an abnormal behaviour, the entire diagnose activity would be triggered for the abnormality presented by that agent, thus delaying the restoration of a normal operation. Therefore, whatever are the remediation actions implemented in a given domain, it is essential to---at least temporarily---ensure their effectiveness in satisfying existing quality feature requirements. How to guarantee such a characteristic is another open research subject.

Finally, our solution also assumes that, when required, agents will always provide the most accurate information based on the data they collect. However, even though we do not address that scenario in this work, we cannot discard the situation in which malicious agents join the system in order to disrupt its behaviour. To handle this limitation, we can take advantage of existing solutions that propose the adoption of different models of reputation~\citep{mui:02,huynh:06} in a way that agents would only take into account information provided by other reliable components. How to extend our solution in order to integrate these approaches is a topic to be investigated.

% !TEX root = main.tex

\section{Related Work}\label{sec:related-work}

Considerable effort has been directed towards the development of solutions to improve the diagnosis, remediation and solution of problem causes in several domains. However, these activities are rarely integrated into single cohesive proposal. In this section, we present related work categorised according to their main goals. %Some of these works may eventually be considered complementary to our solution.

\subsection{Cause Diagnosis}
The concept of Bayesian networks~\citep{pearl:14} is frequently used as a starting point for the development of several approaches that focus on the diagnosis of event causes. In these networks, variables of interest are represented as nodes while causal relationships are depicted as edges. To create such a network, the accountability framework proposed by \cite{zhang:07} maps services and their dependencies to nodes and edges, respectively. A probabilistic analysis is thus performed on the generated structure with the aim of identifying the origin of Service Level Agreement violations. Carrying out this process, however, depends on knowing the topology of service relationships in advance, which hinders the adoption of this solution for dynamic systems. That is the same limitation presented by the solution proposed by~\cite{verbert:17}. Nevertheless, their work purposely targets a sort of systems that is inherently static. They take advantage of Bayesian networks to diagnose the root cause of issues in the context of Heating, Ventilation and Air Conditioning systems. The diagnosis is made according to the conditional probability distribution of faults given a set of observed variables. However, even though these systems having many different components, there is no detail on how such a solution could be implemented in a distributed environment.

The CloudRanger framework~\citep{wang:18} has a similar workflow. At runtime, it creates a network representing the impact services have on others, and uses it as the foundation to compute the correlation between linked services regarding a common cause. Traversing this network according to computed correlations allows one to rank the candidate root causes of a given abnormality. \cite{parida:18}, in turn, proposed a domain-independent solution in which the concept of causal influence factor is introduced as a parameter to be used to identify not only the underlying causal structure of systems but also the directions of existing causal relationships. Despite allowing, the discovery of root causes in dynamic environments with higher accuracy, these proposals do not handle scenarios in which the data are decentralised.

A solution that explores the cooperation between autonomous agents is proposed to handle this issue~\citep{maes:07}. It assumes that agents do not have a global view of the system, and thus are not able to observe the entire set of domain variables. This solution thus proposes a mechanism that allows agents to negotiate the disclosure of useful information, which is later used to create individual causal models and guide adaptation strategies. Even though it targets distributed domains, the feasibility of this approach is not evaluated in scenarios comprising more than two agents. Finally, despite not being designed for diagnosing the root cause of problems, the dynamic, adaptive modeling of the behaviour of cloud-based systems, as proposed by \cite{chen:17}, comprises an interesting alternative to achieve such a goal. Their focus is on predicting the value of Quality of Service (QoS) features based on environmental conditions and other relevant variables. By correct modeling the impact of these variables on QoS features, it becomes possible to identify the cause of quality requirement violations even before their occurrence.

\subsection{Adaptive Strategies}
The need for adaptability in distributed applications has motivated the development of different solutions. A conceptual model named D\textsuperscript{2}R\textsuperscript{2}+DR~\citep{sterbenz:10}, initially designed to promote resilience in the context of networked systems, proposes the integration of activities into two phases. The first is responsible for detecting and mitigating problems, while the second has the goal of diagnosing problem causes and improving the system behaviour in order to prevent further failures. What this model does, in fact, is to provide general guidelines to design systems able to overcome challenging situations. It is clear how our solution relates to both phases described in this model as we provide a concrete solution for mitigating, diagnosing and solving causes of identified abnormalities.

Usually, adaptive strategies aim to improve the system behaviour regarding the satisfaction of specific quality features. Particularly focused on system reliability, \cite{brun:11, brun:15} proposed a strategy based on the redundant execution of tasks, which are assigned to computing components by a centralised orchestrator. The number of components performing the same task is dynamically adapted as a response to component failures with the goal of achieving a predefined reliability level. While in this work component failures are considered independent from each other, in our proposal we explore the existence of such a dependency to diagnose the root cause of problems. To fulfill a range of different QoS requirements, \cite{wang:05} proposed a framework that takes advantage of root cause identification and adaptation to handle different situations. This approach relies on the analysis of a given causal network to identify the cause of QoS contract violations and on the use of if-then decision rules to select predefined adaptation actions that would allow components to deal with challenging scenarios. Fulfilling predefined requirements is also he goal of the approach proposed by \cite{rosa:13}. In this work, they consider components as having multiple implementations, each of them with different, already known impact on existing quality features. The approach comprises offline and online phases. From the knowledge on the impact of different component implementations, the solution is able to specify a set of adaptation rules that can be selected at runtime in order to modify the system to achieve existing quality requirements.

% !TEX root = main.tex

\section{Conclusion}\label{sec:conclusion}

Because of their widespread use and undeniable impact on the achievement of increasingly complex and critical tasks, multiagent systems are required to operate with acceptable quality even in the face of challenging, unforeseen situations. A strategy frequently adopted in order to achieve this goal involves the diagnosis, remediation and solution of problem causes. Activities comprising this strategy, however, rely either on human-in-the-loop or on centralised information and processing.

Motivated by this scenario, in this work we proposed a cooperative strategy particularly focused on the diagnosis of the root causes of service quality requirement violations. This strategy allows autonomous agents to cooperate with each other in order to identify whether these violations come from service providers, associated components, or the communication infrastructure. From this identification process, agents are able to mitigate and repair abnormalities with the aim of normalising system operation. Our cooperative strategy comprises a protocol that provides guidance on how agents should interact as well as the specification of how agents are expected to behave when playing the role of clients, providers and cooperating agents. We remove the need for human involvement by taking advantage of information collected by individual agents at runtime and using it as the source of knowledge to diagnose abnormal behaviour. The centralised information and processing is also relieved when we acknowledge that pieces of information may be spread throughout the system, and thus promote the cooperation among agents. We evaluated our proposal with an empirical experiment in which we simulated a distributed environment comprising 38 autonomous agents. Results indicate that our solution is able to handle a variety of failures at different levels, keeping the system operating with desired quality and reduced costs.

As discussed in Section~\ref{sec:evaluation}, the proposed approach still has its limitations. First, there is a need for cooperating agents to interact with an abnormal component after the occurrence of a quality requirement violation in order to allow the correct diagnosis of the problem cause. Second, it is crucial to ensure the effectiveness of implemented remediation actions regarding the satisfaction of quality requirements. Finally, malicious components joining the system are an issue to be considered. Future work includes the investigation of strategies to handle these limitations.

%Future work includes the investigation of strategies to increase the adaptability of our approach. Currently, abstract operations specified in our solution that are focused on the remediation and solution of identified causes are candidates for the development of more flexible and adaptive implementations. 

\section*{Reproduction Package}
A reproduction package containing the source code used in our evaluation as well as the collected data is available at \url{https://www.inf.ufrgs.br/prosoft/resources/2020/eaai-mas-self-adaptive-protocol}.

\section*{Acknowledgments}
We acknowledge the support of the National Council for Scientific and Technological Development of Brazil (CNPq) (grants ref. 141840/2016-1, 313357/2018-8, 428157/2018-1), and the support of the Government of Canada. This study was financed in part by the Coordena\c{c}\~{a}o de Aperfei\c{c}oamento de Pessoal de N\'{i}vel Superior - Brasil (CAPES) - Finance Code 001.

\bibliographystyle{elsarticle-harv}
\bibliography{bibliography.bib}

\begin{thebibliography}{30}
\expandafter\ifx\csname natexlab\endcsname\relax\def\natexlab#1{#1}\fi
\providecommand{\url}[1]{\texttt{#1}}
\providecommand{\href}[2]{#2}
\providecommand{\path}[1]{#1}
\providecommand{\DOIprefix}{doi:}
\providecommand{\ArXivprefix}{arXiv:}
\providecommand{\URLprefix}{URL: }
\providecommand{\Pubmedprefix}{pmid:}
\providecommand{\doi}[1]{\href{http://dx.doi.org/#1}{\path{#1}}}
\providecommand{\Pubmed}[1]{\href{pmid:#1}{\path{#1}}}
\providecommand{\bibinfo}[2]{#2}
\ifx\xfnm\relax \def\xfnm[#1]{\unskip,\space#1}\fi
%Type = Article
\bibitem[{{Baresi} et~al.(2006){Baresi}, {Di Nitto} and {Ghezzi}}]{baresi:06}
\bibinfo{author}{{Baresi}, L.}, \bibinfo{author}{{Di Nitto}, E.},
  \bibinfo{author}{{Ghezzi}, C.}, \bibinfo{year}{2006}.
\newblock \bibinfo{title}{Toward open-world software: Issues and challenges}.
\newblock \bibinfo{journal}{Computer} \bibinfo{volume}{39},
  \bibinfo{pages}{36--43}.
%Type = Article
\bibitem[{{Brun} et~al.(2015){Brun}, y.~{Bang}, {Edwards} and
  {Medvidovic}}]{brun:15}
\bibinfo{author}{{Brun}, Y.}, \bibinfo{author}{y.~{Bang}, J.},
  \bibinfo{author}{{Edwards}, G.}, \bibinfo{author}{{Medvidovic}, N.},
  \bibinfo{year}{2015}.
\newblock \bibinfo{title}{Self-adapting reliability in distributed software
  systems}.
\newblock \bibinfo{journal}{IEEE Transactions on Software Engineering}
  \bibinfo{volume}{41}, \bibinfo{pages}{764--780}.
\newblock \DOIprefix\doi{10.1109/TSE.2015.2412134}.
%Type = Inproceedings
\bibitem[{{Brun} et~al.(2011){Brun}, {Edwards}, {Bang} and
  {Medvidovic}}]{brun:11}
\bibinfo{author}{{Brun}, Y.}, \bibinfo{author}{{Edwards}, G.},
  \bibinfo{author}{{Bang}, J.Y.}, \bibinfo{author}{{Medvidovic}, N.},
  \bibinfo{year}{2011}.
\newblock \bibinfo{title}{Smart redundancy for distributed computation}, in:
  \bibinfo{booktitle}{2011 31st International Conference on Distributed
  Computing Systems}, pp. \bibinfo{pages}{665--676}.
\newblock \DOIprefix\doi{10.1109/ICDCS.2011.25}.
%Type = Article
\bibitem[{Chandola et~al.(2009)Chandola, Banerjee and Kumar}]{chandola:09}
\bibinfo{author}{Chandola, V.}, \bibinfo{author}{Banerjee, A.},
  \bibinfo{author}{Kumar, V.}, \bibinfo{year}{2009}.
\newblock \bibinfo{title}{Anomaly detection: A survey}.
\newblock \bibinfo{journal}{ACM Computing Surveys} \bibinfo{volume}{41}.
\newblock \DOIprefix\doi{10.1145/1541880.1541882}.
%Type = Article
\bibitem[{{Chen} and {Bahsoon}(2017)}]{chen:17}
\bibinfo{author}{{Chen}, T.}, \bibinfo{author}{{Bahsoon}, R.},
  \bibinfo{year}{2017}.
\newblock \bibinfo{title}{Self-adaptive and online {QoS} modeling for
  cloud-based software services}.
\newblock \bibinfo{journal}{IEEE Transactions on Software Engineering}
  \bibinfo{volume}{43}, \bibinfo{pages}{453--475}.
\newblock \DOIprefix\doi{10.1109/TSE.2016.2608826}.
%Type = Article
\bibitem[{{De Sanctis} et~al.(2020){De Sanctis}, {Bucchiarone} and
  {Marconi}}]{deSanctis:20}
\bibinfo{author}{{De Sanctis}, M.}, \bibinfo{author}{{Bucchiarone}, A.},
  \bibinfo{author}{{Marconi}, A.}, \bibinfo{year}{2020}.
\newblock \bibinfo{title}{Dynamic adaptation of service-based applications: a
  design for adaptation approach}.
\newblock \bibinfo{journal}{Journal of Internet Services and Applications}
  \bibinfo{volume}{11}.
\newblock \DOIprefix\doi{10.1186/s13174-020-00123-6}.
%Type = Article
\bibitem[{Dobson et~al.(2006)Dobson, Denazis, Fern\'{a}ndez, Ga\"{\i}ti,
  Gelenbe, Massacci, Nixon, Saffre, Schmidt and Zambonelli}]{dobson:06}
\bibinfo{author}{Dobson, S.}, \bibinfo{author}{Denazis, S.},
  \bibinfo{author}{Fern\'{a}ndez, A.}, \bibinfo{author}{Ga\"{\i}ti, D.},
  \bibinfo{author}{Gelenbe, E.}, \bibinfo{author}{Massacci, F.},
  \bibinfo{author}{Nixon, P.}, \bibinfo{author}{Saffre, F.},
  \bibinfo{author}{Schmidt, N.}, \bibinfo{author}{Zambonelli, F.},
  \bibinfo{year}{2006}.
\newblock \bibinfo{title}{A survey of autonomic communications}.
\newblock \bibinfo{journal}{ACM Transactions on Autonomous and Adaptive
  Systems} \bibinfo{volume}{1}, \bibinfo{pages}{223--259}.
\newblock \DOIprefix\doi{10.1145/1186778.1186782}.
%Type = Article
\bibitem[{{Dobson} et~al.(2019){Dobson}, {Hutchison}, {Mauthe},
  {Schaeffer-Filho}, {Smith} and {Sterbenz}}]{dobson:19}
\bibinfo{author}{{Dobson}, S.}, \bibinfo{author}{{Hutchison}, D.},
  \bibinfo{author}{{Mauthe}, A.}, \bibinfo{author}{{Schaeffer-Filho}, A.},
  \bibinfo{author}{{Smith}, P.}, \bibinfo{author}{{Sterbenz}, J.P.G.},
  \bibinfo{year}{2019}.
\newblock \bibinfo{title}{Self-organization and resilience for networked
  systems: Design principles and open research issues}.
\newblock \bibinfo{journal}{Proceedings of the IEEE} \bibinfo{volume}{107},
  \bibinfo{pages}{819--834}.
\newblock \DOIprefix\doi{10.1109/JPROC.2019.2894512}.
%Type = Inproceedings
\bibitem[{D{\"o}tterl et~al.(2017)D{\"o}tterl, Bruns, Dunkel and
  Ossowski}]{Dotterl:17}
\bibinfo{author}{D{\"o}tterl, J.}, \bibinfo{author}{Bruns, R.},
  \bibinfo{author}{Dunkel, J.}, \bibinfo{author}{Ossowski, S.},
  \bibinfo{year}{2017}.
\newblock \bibinfo{title}{Towards dynamic rebalancing of bike sharing systems:
  An event-driven agents approach}, in: \bibinfo{editor}{Oliveira, E.},
  \bibinfo{editor}{Gama, J.}, \bibinfo{editor}{Vale, Z.},
  \bibinfo{editor}{Lopes~Cardoso, H.} (Eds.), \bibinfo{booktitle}{Progress in
  Artificial Intelligence}, \bibinfo{publisher}{Springer International
  Publishing}, \bibinfo{address}{Cham}. pp. \bibinfo{pages}{309--320}.
%Type = Inproceedings
\bibitem[{Faccin and Nunes(2018a)}]{faccin:18b}
\bibinfo{author}{Faccin, J.}, \bibinfo{author}{Nunes, I.},
  \bibinfo{year}{2018}a.
\newblock \bibinfo{title}{Cleaning up the mess: A formal framework for
  autonomously reverting bdi agent actions}, in:
  \bibinfo{booktitle}{Proceedings of the 13th International Conference on
  Software Engineering for Adaptive and Self-Managing Systems},
  \bibinfo{publisher}{Association for Computing Machinery},
  \bibinfo{address}{New York, NY, USA}. p. \bibinfo{pages}{108–118}.
\newblock \DOIprefix\doi{10.1145/3194133.3194156}.
%Type = Article
\bibitem[{Faccin and Nunes(2018b)}]{faccin:18a}
\bibinfo{author}{Faccin, J.}, \bibinfo{author}{Nunes, I.},
  \bibinfo{year}{2018}b.
\newblock \bibinfo{title}{Remediating critical cause-effect situations with an
  extended {BDI} architecture}.
\newblock \bibinfo{journal}{Expert Systems with Applications}
  \bibinfo{volume}{95}, \bibinfo{pages}{190--200}.
\newblock \DOIprefix\doi{10.1016/j.eswa.2017.11.036}.
%Type = Techreport
\bibitem[{{Foundation for Intelligent Physical Agents}(2002)}]{fipa:02}
\bibinfo{author}{{Foundation for Intelligent Physical Agents}},
  \bibinfo{year}{2002}.
\newblock \bibinfo{title}{{FIPA} Contract Net Interaction Protocol
  Specification}.
\newblock \bibinfo{type}{Technical Report}.
\newblock \URLprefix \url{http://www.fipa.org/specs/fipa00029/SC00029H.html}.
%Type = Article
\bibitem[{Ghosh et~al.(2007)Ghosh, Sharman, Rao and Upadhyaya}]{ghosh:07}
\bibinfo{author}{Ghosh, D.}, \bibinfo{author}{Sharman, R.},
  \bibinfo{author}{Rao, H.R.}, \bibinfo{author}{Upadhyaya, S.},
  \bibinfo{year}{2007}.
\newblock \bibinfo{title}{Self-healing systems --- survey and synthesis}.
\newblock \bibinfo{journal}{Decision Support Systems} \bibinfo{volume}{42},
  \bibinfo{pages}{2164--2185}.
\newblock \DOIprefix\doi{10.1016/j.dss.2006.06.011}.
%Type = Inproceedings
\bibitem[{Huang and Nitschke(2020)}]{Huang:20}
\bibinfo{author}{Huang, A.}, \bibinfo{author}{Nitschke, G.},
  \bibinfo{year}{2020}.
\newblock \bibinfo{title}{Automating coordinated autonomous vehicle control},
  in: \bibinfo{booktitle}{Proceedings of the 19th International Conference on
  Autonomous Agents and MultiAgent Systems}, \bibinfo{publisher}{International
  Foundation for Autonomous Agents and Multiagent Systems},
  \bibinfo{address}{Richland, SC}. pp. \bibinfo{pages}{1867--1868}.
%Type = Article
\bibitem[{Huynh et~al.(2006)Huynh, Jennings and Shadbolt}]{huynh:06}
\bibinfo{author}{Huynh, T.D.}, \bibinfo{author}{Jennings, N.R.},
  \bibinfo{author}{Shadbolt, N.R.}, \bibinfo{year}{2006}.
\newblock \bibinfo{title}{An integrated trust and reputation model for open
  multi-agent systems}.
\newblock \bibinfo{journal}{Autonomous Agents and Multi-Agent Systems}
  \bibinfo{volume}{13}, \bibinfo{pages}{119--154}.
\newblock \DOIprefix\doi{10.1007/s10458-005-6825-4}.
%Type = Inproceedings
\bibitem[{{Insaurralde}(2014)}]{Insaurralde:14}
\bibinfo{author}{{Insaurralde}, C.C.}, \bibinfo{year}{2014}.
\newblock \bibinfo{title}{Service-oriented agent architecture for unmanned air
  vehicles}, in: \bibinfo{booktitle}{IEEE/AIAA Digital Avionics Systems
  Conference (DASC)}, pp. \bibinfo{pages}{1--19}.
%Type = Article
\bibitem[{Jennings(2001)}]{Jennings:01}
\bibinfo{author}{Jennings, N.R.}, \bibinfo{year}{2001}.
\newblock \bibinfo{title}{An agent-based approach for building complex software
  systems}.
\newblock \bibinfo{journal}{Commun. ACM} \bibinfo{volume}{44},
  \bibinfo{pages}{35–--41}.
\newblock \DOIprefix\doi{10.1145/367211.367250}.
%Type = Article
\bibitem[{Maes et~al.(2007)Maes, Meganck and Manderick}]{maes:07}
\bibinfo{author}{Maes, S.}, \bibinfo{author}{Meganck, S.},
  \bibinfo{author}{Manderick, B.}, \bibinfo{year}{2007}.
\newblock \bibinfo{title}{Inference in multi-agent causal models}.
\newblock \bibinfo{journal}{International Journal of Approximate Reasoning}
  \bibinfo{volume}{46}, \bibinfo{pages}{274--299}.
\newblock \DOIprefix\doi{10.1016/j.ijar.2006.09.005}.
%Type = Inproceedings
\bibitem[{Mui et~al.(2002)Mui, Mohtashemi and Halberstadt}]{mui:02}
\bibinfo{author}{Mui, L.}, \bibinfo{author}{Mohtashemi, M.},
  \bibinfo{author}{Halberstadt, A.}, \bibinfo{year}{2002}.
\newblock \bibinfo{title}{Notions of reputation in multi-agents systems: A
  review}, in: \bibinfo{booktitle}{Proceedings of the First International Joint
  Conference on Autonomous Agents and Multiagent Systems},
  \bibinfo{publisher}{Association for Computing Machinery},
  \bibinfo{address}{New York, NY, USA}. pp. \bibinfo{pages}{280--287}.
\newblock \DOIprefix\doi{10.1145/544741.544807}.
%Type = Article
\bibitem[{Parida et~al.(2018)Parida, Marwala and Chakraverty}]{parida:18}
\bibinfo{author}{Parida, P.K.}, \bibinfo{author}{Marwala, T.},
  \bibinfo{author}{Chakraverty, S.}, \bibinfo{year}{2018}.
\newblock \bibinfo{title}{A multivariate additive noise model for complete
  causal discovery}.
\newblock \bibinfo{journal}{Neural Networks} \bibinfo{volume}{103},
  \bibinfo{pages}{44--54}.
\newblock \DOIprefix\doi{10.1016/j.neunet.2018.03.013}.
%Type = Article
\bibitem[{Parzen(1962)}]{parzen:62}
\bibinfo{author}{Parzen, E.}, \bibinfo{year}{1962}.
\newblock \bibinfo{title}{On estimation of a probability density function and
  mode}.
\newblock \bibinfo{journal}{Annals of Mathematical Statistics}
  \bibinfo{volume}{33}, \bibinfo{pages}{1065--1076}.
\newblock \DOIprefix\doi{10.1214/aoms/1177704472}.
%Type = Book
\bibitem[{Pearl(2014)}]{pearl:14}
\bibinfo{author}{Pearl, J.}, \bibinfo{year}{2014}.
\newblock \bibinfo{title}{Probabilistic reasoning in intelligent systems:
  networks of plausible inference}.
\newblock \bibinfo{publisher}{Elsevier}.
%Type = Article
\bibitem[{{Rosa} et~al.(2013){Rosa}, {Rodrigues}, {Lopes}, {Hiltunen} and
  {Schlichting}}]{rosa:13}
\bibinfo{author}{{Rosa}, L.}, \bibinfo{author}{{Rodrigues}, L.},
  \bibinfo{author}{{Lopes}, A.}, \bibinfo{author}{{Hiltunen}, M.},
  \bibinfo{author}{{Schlichting}, R.}, \bibinfo{year}{2013}.
\newblock \bibinfo{title}{Self-management of adaptable component-based
  applications}.
\newblock \bibinfo{journal}{IEEE Transactions on Software Engineering}
  \bibinfo{volume}{39}, \bibinfo{pages}{403--421}.
\newblock \DOIprefix\doi{10.1109/TSE.2012.29}.
%Type = Article
\bibitem[{Sterbenz et~al.(2010)Sterbenz, Hutchison, Çetinkaya, Jabbar, Rohrer,
  Schöller and Smith}]{sterbenz:10}
\bibinfo{author}{Sterbenz, J.P.}, \bibinfo{author}{Hutchison, D.},
  \bibinfo{author}{Çetinkaya, E.K.}, \bibinfo{author}{Jabbar, A.},
  \bibinfo{author}{Rohrer, J.P.}, \bibinfo{author}{Schöller, M.},
  \bibinfo{author}{Smith, P.}, \bibinfo{year}{2010}.
\newblock \bibinfo{title}{Resilience and survivability in communication
  networks: Strategies, principles, and survey of disciplines}.
\newblock \bibinfo{journal}{Computer Networks} \bibinfo{volume}{54},
  \bibinfo{pages}{1245--1265}.
\newblock \DOIprefix\doi{10.1016/j.comnet.2010.03.005}.
%Type = Book
\bibitem[{Tukey(1977)}]{tukey:77}
\bibinfo{author}{Tukey, J.W.}, \bibinfo{year}{1977}.
\newblock \bibinfo{title}{Exploratory Data Analysis}.
\newblock \bibinfo{publisher}{Addison-Wesley}.
%Type = Article
\bibitem[{Verbert et~al.(2017)Verbert, Babuška and Schutter}]{verbert:17}
\bibinfo{author}{Verbert, K.}, \bibinfo{author}{Babuška, R.},
  \bibinfo{author}{Schutter, B.D.}, \bibinfo{year}{2017}.
\newblock \bibinfo{title}{Combining knowledge and historical data for
  system-level fault diagnosis of {HVAC} systems}.
\newblock \bibinfo{journal}{Engineering Applications of Artificial
  Intelligence} \bibinfo{volume}{59}, \bibinfo{pages}{260--273}.
\newblock \DOIprefix\doi{10.1016/j.engappai.2016.12.021}.
%Type = Inproceedings
\bibitem[{Wang et~al.(2005)Wang, Wang, Chen, Wang, Fung, Uczekaj, Chen,
  Guthmiller and Lee}]{wang:05}
\bibinfo{author}{Wang, G.}, \bibinfo{author}{Wang, C.}, \bibinfo{author}{Chen,
  A.}, \bibinfo{author}{Wang, H.}, \bibinfo{author}{Fung, C.},
  \bibinfo{author}{Uczekaj, S.}, \bibinfo{author}{Chen, Y.L.},
  \bibinfo{author}{Guthmiller, W.}, \bibinfo{author}{Lee, J.},
  \bibinfo{year}{2005}.
\newblock \bibinfo{title}{Service level management using {QoS} monitoring,
  diagnostics, and adaptation for networked enterprise systems}, in:
  \bibinfo{booktitle}{Ninth IEEE International EDOC Enterprise Computing
  Conference (EDOC'05)}, pp. \bibinfo{pages}{239--248}.
\newblock \DOIprefix\doi{10.1109/EDOC.2005.30}.
%Type = Inproceedings
\bibitem[{{Wang} et~al.(2018){Wang}, {Xu}, {Ma}, {Lin}, {Pan}, {Wang} and
  {Chen}}]{wang:18}
\bibinfo{author}{{Wang}, P.}, \bibinfo{author}{{Xu}, J.},
  \bibinfo{author}{{Ma}, M.}, \bibinfo{author}{{Lin}, W.},
  \bibinfo{author}{{Pan}, D.}, \bibinfo{author}{{Wang}, Y.},
  \bibinfo{author}{{Chen}, P.}, \bibinfo{year}{2018}.
\newblock \bibinfo{title}{Cloudranger: Root cause identification for cloud
  native systems}, in: \bibinfo{booktitle}{2018 18th IEEE/ACM International
  Symposium on Cluster, Cloud and Grid Computing (CCGRID)}, pp.
  \bibinfo{pages}{492--502}.
\newblock \DOIprefix\doi{10.1109/CCGRID.2018.00076}.
%Type = Article
\bibitem[{Zhang et~al.(2007)Zhang, Lin and Hsu}]{zhang:07}
\bibinfo{author}{Zhang, Y.}, \bibinfo{author}{Lin, K.J.}, \bibinfo{author}{Hsu,
  J.Y.J.}, \bibinfo{year}{2007}.
\newblock \bibinfo{title}{Accountability monitoring and reasoning in
  service-oriented architectures}.
\newblock \bibinfo{journal}{Service Oriented Computing and Applications}
  \bibinfo{volume}{1}, \bibinfo{pages}{35--50}.
\newblock \DOIprefix\doi{10.1007/s11761-007-0001-4}.
%Type = Article
\bibitem[{{Zhou} et~al.(2018){Zhou}, {Peng}, {Xie}, {Sun}, {Ji}, {Li} and
  {Ding}}]{zhou:18}
\bibinfo{author}{{Zhou}, X.}, \bibinfo{author}{{Peng}, X.},
  \bibinfo{author}{{Xie}, T.}, \bibinfo{author}{{Sun}, J.},
  \bibinfo{author}{{Ji}, C.}, \bibinfo{author}{{Li}, W.},
  \bibinfo{author}{{Ding}, D.}, \bibinfo{year}{2018}.
\newblock \bibinfo{title}{Fault analysis and debugging of microservice systems:
  Industrial survey, benchmark system, and empirical study}.
\newblock \bibinfo{journal}{IEEE Transactions on Software Engineering}
  \bibinfo{volume}{14}, \bibinfo{pages}{1--18}.

\end{thebibliography}

\end{document}